\documentclass[pra,twocolumn,floatfix,superscriptaddress]{revtex4-1}
\usepackage{graphicx}
\usepackage{ulem}
\usepackage{multirow}
\usepackage{amsmath, calc}
\usepackage{hyperref}
\usepackage{bm}

\graphicspath{ {./images/} }

\newcommand{\WaxeN}{W_{\textrm{ax}}^{(eN)}}
\newcommand{\Waxee}{W_{\textrm{ax}}^{(ee)}}
\newcommand{\Ti}{$\mathcal{T}$}
\newcommand{\Par}{$\mathcal{P}$}
\newcommand{\CP}{$\mathcal{CP}$}

\newcommand{\de}{d_\mathrm{e}}

\usepackage{color}

\usepackage{hyperref}
\hypersetup{colorlinks=true, citecolor=blue, urlcolor=blue, linkcolor=blue}

\date{02.04.2024}

\begin{abstract}
The effects induced by the time-reversal (\Ti) and spatial parity (\Par) violating electron-nucleus and electron-electron interactions mediated by the axion-like particles (ALPs) in the BaF molecule were estimated. Molecular parameters characterizing these interactions were calculated across a wide range of ALP masses. In case of the electron-nuclear interaction, the effect of the extended nucleus was studied and shown to be significant for heavy ALPs. Based on the calculated molecular parameters, we obtained a link between the \Ti,\Par-violating energy shift which can be measured in the ongoing experiment designed to search for the electron electric dipole moment using the BaF molecule [A. Boeschoten et al., arXiv:2303.06402 (2023)] and the products of coupling constants of the ALP-electron or ALP-electron and ALP-nucleus interactions.
\end{abstract}

\begin{document}

\title{Axion-mediated electron-nucleus and electron-electron interactions in barium monofluoride molecule}

\author{Sergey D.\ Prosnyak}
\email{prosnyak\_sd@pnpi.nrcki.ru, prosnyak.sergey@yandex.ru}
\affiliation{Petersburg Nuclear Physics Institute named by B.P. Konstantinov of National Research Centre
``Kurchatov Institute'', Gatchina, Leningrad District 188300, Russia}
\affiliation{Saint Petersburg State University, 7/9 Universitetskaya nab., St. Petersburg, 199034 Russia}
\author{Leonid V.\ Skripnikov}
\email{skripnikov\_lv@pnpi.nrcki.ru,\\ leonidos239@gmail.com}
\homepage{http://www.qchem.pnpi.spb.ru}
\affiliation{Petersburg Nuclear Physics Institute named by B.P. Konstantinov of National Research Centre
``Kurchatov Institute'', Gatchina, Leningrad District 188300, Russia}
\affiliation{Saint Petersburg State University, 7/9 Universitetskaya nab., St. Petersburg, 199034 Russia}

\maketitle
 
\section{Introduction}

Precise experiments with heavy-atom molecules, aimed at detecting violations of spatial parity (\Par) and time-reversal (\Ti) symmetries in fundamental interactions, are considered one of the most powerful tools for exploring new physics beyond the Standard Model~(SM)~\cite{Safronova:18,eEDM_snowmass:2022,Safronova2023}. The permanent \Ti,\Par-violating electric dipole moments (EDM) of atoms and molecules can be induced by the permanent electron electric dipole moment ($e$EDM or $\de$). New physics extensions of the Standard Model predict the $e$EDM value to be orders of magnitude larger than the estimates within the SM~\cite{eEDMLimit:2022,Yamaguchi:2021,Engel:2013,Chubukov:2016,Commins:98,Chupp:2019}. 
Recently, the JILA group updated the most stringent upper bound on the $e$EDM through experiments with hafnium monofluoride (HfF$^+$) molecular cations, establishing a 90\% confidence level of $|\de| < 4.1 \times 10^{-30}$ $\ensuremath{e \cdot {\rm cm}}$~\cite{newlimit1}. This updated constraint improves upon the previous most stringent limit obtained using the ThO molecular beam by a factor of 2.4~\cite{ACME:18}.

In Ref.~\cite{BaF:2018}, an experiment was proposed to search for the $e$EDM using an intense cold beam of barium monofluoride molecules. The expected sensitivity to the $e$EDM was estimated as $|\de|= 5.0\times 10^{-30}$ $\ensuremath{e \cdot {\rm cm}}$~\cite{BaF:2018}. This level of sensitivity is comparable to that achieved in the HfF$^+$ experiment. Such precision can be attained by employing a combination of Stark deceleration, laser cooling, and an intense primary cold source of BaF molecules. In Ref.~\cite{boeschoten2023novel}, a new method was proposed to reduce systematic errors and to reduce the need for numerous auxiliary measurements.
A further step towards laser cooling of ${}^{138}$Ba${}^{19}$F molecules that takes into account the hyperfine splitting has been recently made in Ref.~\cite{PhysRevA.108.062812}.

In addition to the effect of \Ti,\Par-violation in atoms and molecules induced by the $e$EDM, the scalar-pseudoscalar nucleus-electron interaction is often considered~\cite{KL95, Ginges:04, ACME:18, newlimit1}. However, other mechanisms can also induce such effects. Here, we consider the \Ti,\Par-violating interactions arising from the exchange of a virtual axion-like particle (ALP) between the electron and nucleus, as well as between electrons in the BaF molecule. Axions and ALPs are popular dark matter candidates~\cite{abbott1983cosmological, preskill1983cosmology, dine1983not}. Axion appears in the Peccei-Quinn solution to the strong \CP\ problem~\cite{peccei1977cp, wilczek1978problem, weinberg1978new} and String Theory compactifications~\cite{svrcek2006axions, arvanitaki2010string}. Therefore, these particles get an increasing attention from both theory and experiment. Numerous astrophysical and laboratory experiments are devoted to search for axions and ALPs and study their properties. A compilation of the constraints on axion and ALP properties obtained so far can be found e.g. in Ref.~\cite{Hare:2020}. Recently, there have been proposals for new atomic and molecular experiments aimed at searching for axion-induced forces using spin-based quantum sensors, see e.g. Refs.~\cite{agrawal2023searching, kimball2023probing} and references therein.

In Ref.~\cite{Stadnik:2018}, direct calculations were performed to determine the \Ti,\Par-violating effects mediated by ALPs on electron-nucleus and electron-electron interactions in various atoms, with pilot estimates provided for some molecules. The \textit{ab-initio} studies of this effect for the YbOH molecule, and HfF$^+$ molecular cation as well as for the Fr atom were carried out in Refs.~\cite{Maison:2021,maison2021axion,Maison:2022,Prosnyak:2023a}.

In the present paper, we estimated the \Ti,\Par-violating effects arising from ALP exchange between electrons and between the electron and nucleus for the ${}^{138}$Ba${}^{19}$F molecule in its ground electronic state $X^2\Sigma_{1/2}$. We employed the theoretical approach developed in Ref.~\cite{Prosnyak:2023a}, enabling the direct calculation of the contribution of these effects in molecules across a wide range of ALP masses. In this work, we extended this approach to include the finite-nucleus size effect in the case of the nucleus-electron interaction.

\section{Theory}

The \Ti,\Par-violating interaction between an electron and a nucleon, mediated by an ALP with a mass of $m_a$, can be expressed in the following form~\cite{moody1984new, Stadnik:2018}:
\begin{equation}
\label{eN_potential}
    V_{eN} (\boldsymbol{r}-\boldsymbol{R}) = +i \frac{g_N^sg_e^p}{4 \pi}
    \frac{e^{-m_a |\boldsymbol{r} - \boldsymbol{R}|}}{|\boldsymbol{r} - \boldsymbol{R}|} \gamma_0 \gamma_5,
\end{equation}
where $g_N^s$ is the coupling constant of the scalar ALP-nucleon interaction,  $g_e^p$ represents the coupling constant of the pseudoscalar ALP-electron interaction, $\boldsymbol{r}$ is the position of the electron, $\boldsymbol{R}$ is the position of the nucleon, and Dirac $\gamma$ matrices refer to the electron, $N$ is the nucleon, which can be either a proton or a neutron. 
In Ref.~\cite{Prosnyak:2023a}, which focused on the estimation of the ALP-induced \Ti,\Par-violating effect in HfF$^+$, we approximated the nucleus by a point-like particle in the \Ti,\Par-violating interaction calculation. However, it is worth noting that the extended  nuclear charge distribution was considered in the calculation of the electronic wave function. Thus, the following approximation for the \Ti,\Par-violating electron-nucleus interaction was employed:
\begin{equation}
\label{eN_point_potential}
    V_{eN} (\boldsymbol{r}) = +i \frac{g_N^sg_e^p}{4 \pi}
    \frac{e^{-m_a r}}{r} \gamma_0 \gamma_5.
\end{equation}

Another \Ti,\Par-violation effect is induced by the ALP-mediated interaction between two electrons, which can be expressed as follows~\cite{moody1984new,Stadnik:2018}:
\begin{equation} 
\label{ee_potential}
    V_{ee}(\boldsymbol{r}_1, \boldsymbol{r}_2) = 
    +i\frac{g_e^s g_e^p}{4\pi} \frac{e^{-m_a |\boldsymbol{r}_1 - \boldsymbol{r}_2|}}{|\boldsymbol{r}_1 - \boldsymbol{r}_2|} \gamma_0 \gamma_5 \ ,
\end{equation}
where $g_e^s$ represents the scalar coupling constant of the electron-ALP interaction, $\boldsymbol{r}_1$ and $\boldsymbol{r}_2$ are the positions of the electrons, $\gamma_0$ and $\gamma_5$ matrices correspond to the second electron.

The \Ti,\Par-violating interaction between electrons and nucleus under consideration can be characterized by a parameter that is determined by the molecular electronic structure~\cite{Maison:2021,Dmitriev:92}:
\begin{eqnarray}
     \WaxeN(m_a)  =    \nonumber \\
\label{WaxeNME}        
    \frac{1}{\Omega}
    \frac{1}{\bar{g}_N^s g_e^p}
    \langle \Psi | \sum\limits_{i=1}^{N_\textrm{e}} \sum\limits_{N} \int d\boldsymbol{R} 
    \rho(\boldsymbol{R})
    V_{eN}(\boldsymbol{r}_i-\boldsymbol{R}) | \Psi \rangle,    
\end{eqnarray}
where $\Psi$ is the molecular electronic wave function, $\Omega$ is the projection of the total electronic angular momentum on the molecular axis, $N_e$ is the number of electrons, index $i$ runs over all electrons, 
$\rho(\boldsymbol{R})$ is the nuclear density, normalized to unity \footnote{It is assumed that the distributions of neutrons and protons are equal.}.
$\bar{g}_{N}^s$ is the ALP-nucleon coupling constant averaged over the nucleons of the Ba nucleus: $\bar{g}_{N}^s = (N_n g_n^s+Zg_p^s)/A$, $g_n^s$ and $g_p^s$ are the ALP-neutron and ALP-proton coupling constants, respectively, $Z$ is the charge of the Ba nucleus, $N_n$ is its neutron number, and $A=Z+N_n$ with $Z=56$, $N_n=82$, and $A=138$ for the considered case of $^{138}$Ba. 

Similarly, the electron-electron interaction can be described by the following molecular parameter:
\begin{equation}
    \Waxee(m_a) = \frac{1}{\Omega} \frac{1}{g_e^s g_e^p}
    \langle \Psi | 
    \mathop{{\sum}'}_{i,j=1}^{N_e} V_{ee}(\boldsymbol{r}_i, \boldsymbol{r}_j) | \Psi\rangle.
\label{WaxeeME}    
\end{equation}
The prime index in the sum indicates that terms with $i = j$ should be excluded.

The energy shift of the electronic level caused by the \Ti,\Par-violating interaction~(\ref{eN_potential}) can be expressed as follows:
\begin{equation}
\label{Etp_eN}
    \delta E = \bar{g}_{N}^s g_e^p  \Omega \WaxeN(m_a),
\end{equation}
while the energy shift induced by the \Ti,\Par-violating interaction~(\ref{ee_potential}) can be expressed as:
\begin{equation}
\label{Etp_ee}
    \delta E = g_e^s g_e^p \Omega \Waxee(m_a).
\end{equation}

\section{Computational Details}

In Ref.~\cite{Prosnyak:2023a}, the method to calculate the primitive integrals $\langle ab|e^{-m_a r}/r|cd\rangle$ over Gaussian-type basis functions $a,b,c,d$ of the form $x^n y^m z^k e^{-\beta r^2}$, used in the decomposition of molecular bispinors was implemented. The algorithm was specifically applied to calculate molecular parameters $\Waxee(m_a)$ for the molecular cation HfF${}^+$. In this study, we adapted this method 
to calculate molecular parameter $\WaxeN(m_a)$, taking into account the effects of the finite nuclear size. We considered a model of the Gaussian nucleons distribution, which has the same form as the Gaussian nuclear charge distribution model $\rho = c e^{- \alpha r^2}$~\cite{Visscher:1997}. By calculating $\langle a \gamma|e^{-m_a r}/r|c\gamma \rangle$, where $\gamma = \sqrt{\rho}$,
we obtained primitive integrals required to calculate the molecular parameter $\WaxeN(m_a)$ using the extended nucleus model. To calculate the $\Waxee(m_a)$ parameter, we used the approach developed in Ref.~\cite{Prosnyak:2023a}.

The molecular parameters $\WaxeN(m_a)$ and $\Waxee(m_a)$ can be non-zero for electronic states with unpaired electrons. In the ground electronic state $X^2\Sigma_{1/2}$ of BaF, the unpaired electron is primarily localized on the Ba atom and corresponds approximately to the $6s$ state of the Ba$^+$ ion. 
Therefore, in calculations, we neglected the contribution from the F nucleus. According to our estimations, the largest effect of this type, only about 1\%, is achieved for light ALPs and decreases with an increase of the ALPs mass.

For high ALP masses, the contribution from two-center integrals to the molecular parameter $\Waxee(m_a)$ can be expected to be very small due to the short range of the interaction described by Eq.~(\ref{ee_potential}). Indeed, according to our direct 
estimate, starting from the ALP mass of $10^4$~eV, the contributions from the two-center integrals become negligible. As a result, the computation procedure can be simplified by considering only one-center integrals. We followed this approach in calculations of the molecular parameter $\Waxee(m_a)$ for ALP masses greater than $10^4$~eV.

Electronic structure calculations were performed using the locally modified versions of {\sc dirac}~\cite{DIRAC19,Saue:2020} and {\sc mrcc}~\cite{MRCC2020,Kallay:1,Kallay:2} codes within the relativistic coupled cluster method with single, double, and perturbative triple excitation amplitudes, CCSD(T)~\cite{Visscher:96a,Bartlett:2007} to treat electron correlation effects. The Dirac-Coulomb Hamiltonian was employed. For the ground electronic state $X^2\Sigma_{1/2}$ of the BaF molecule, we used an inter-atomic distance value of $R_e = 2.16 \text{\AA}$, closely matching the experimental value~\cite{Huber:1979, Ryzlewicz:1980}. All electrons of the BaF molecule were included in the electron correlation treatment, and a virtual orbitals cutoff was set to 10,000 Hartree. The choice of such a large cutoff is crucial to properly consider the correlation contributions of the inner-core electrons in systems with heavy atoms~\cite{Skripnikov:17a,Skripnikov:15b}. Following Ref.~\cite{Prosnyak:2023a}, we used the finite-field approach to calculate  molecular parameters~$\WaxeN(m_a)$ and ~$\Waxee(m_a)$. The uncontracted Dyall's AE3Z~\cite{Dyall:2016} basis set was used for F, and an extended version of the AE3Z~\cite{dyall2009relativistic, Dyall:12} basis set was used for Ba. In the latter case we added several diffuse basis functions, and in total the basis set centered on Ba consists of $33s$-, $26p$-, $18d$-, $8f$- and $3g$- type functions. Due to the complexity of calculations involving two-electron operators, the AE2Z~\cite{dyall2009relativistic,Dyall:12,Dyall:2016} basis set was employed for both atoms to calculate the molecular parameter $\Waxee(m_a)$. In all-electron calculations, the Gaussian nuclear charge distribution model, well-suited for molecular problems~\cite{Visscher:1997} was used with root-mean-square charge radius calculated by the interpolation formula from Ref.~\cite{Johnson:1985}.

\section{Results and Discussion}

To assess the impact of the extended nucleus size effect on the $\WaxeN(m_a)$ molecular parameter, we first considered the HfF$^+$ molecular cation. Previously~\cite{Prosnyak:2023a}, we employed the point nucleus model approximation (\ref{eN_point_potential}) corresponding to the $\delta$-type distribution in Eq.~(\ref{WaxeNME}). We compared those values of $\WaxeN(m_a)$ for the various values of $m_a$ with the values $\WaxeN(m_a)$ calculated for the nucleon distribution function in Eq.~(\ref{WaxeNME}), approximated by the Gaussian distribution model with the rms radius equal to the  nuclear rms charge radius. We found that the finite nuclear size effect is negligible for 
ALPs with $m_a < 10^7$~eV. 
A small effect appears for $m_a=10^7$ eV
and becomes significant for heavier ALPs
(see Tables I and II in Supplementary Material).
It reaches 15\% for $m_a = 10^{10}$ eV and stabilizes.
These findings can be explained as follows. The characteristic radius of the Yukawa-type interaction $R_{\textrm{Yu}} = 1/m_a (\textrm{relativistic units}) = \hbar/m_a c$ is significantly larger than the molecule size for light ALPs~\cite{Stadnik:2018}. In particular, the characteristic atomic scale of $1$ Bohr corresponds to $m_a\approx 4$~keV. Thus, the details of the nucleons distribution are insignificant for such an interaction. In contrast, for heavier ALPs the characteristic radius can become compatible with the nuclear size. For example, the nuclear scale of $1$~Fermi corresponds to $m_a \approx 0.2$~GeV. Therefore, the effect of the finite size of the atomic nucleus should be taken into account.

\begin{table}[h]
    \caption{The calculated values of $\WaxeN(m_a)$ molecular parameters (in units of $m_e c / \hbar$) obtained at different levels of electronic structure theory.}
    \centering
    \begin{tabular}{lccc}
    \hline
    \hline
    $m_a$, eV  & DHF & CCSD & CCSD(T) \\
    \hline
    1 & $+1.41 \cdot 10^{-5}$& $+1.77 \cdot 10^{-5}$& $+1.74 \cdot 10^{-5}$\\
    10 & $+1.41 \cdot 10^{-5}$& $+1.77 \cdot 10^{-5}$& $+1.74 \cdot 10^{-5}$\\
    10$^2$ & $+1.40 \cdot 10^{-5}$& $+1.77 \cdot 10^{-5}$& $+1.73 \cdot 10^{-5}$\\
    10$^3$ & $+1.11 \cdot 10^{-5}$& $+1.48 \cdot 10^{-5}$& $+1.45 \cdot 10^{-5}$\\
    10$^4$ & $+1.27 \cdot 10^{-6}$& $+1.89 \cdot 10^{-6}$& $+1.86 \cdot 10^{-6}$\\
    10$^5$ & $-6.91\cdot 10^{-6}$& $-1.09\cdot 10^{-5}$& $-1.07\cdot 10^{-5}$\\
    10$^6$ & $-3.19\cdot 10^{-6}$& $-4.79\cdot 10^{-6}$& $-4.70\cdot 10^{-6}$\\
    10$^7$ & $-9.20\cdot 10^{-8}$& $-1.38\cdot 10^{-7}$& $-1.36\cdot 10^{-7}$\\
    10$^8$ & $-1.29 \cdot 10^{-9}$& $-1.94 \cdot 10^{-9}$& $-1.91 \cdot 10^{-9}$\\
    10$^9$ & $-1.34 \cdot 10^{-11}$& $-2.01 \cdot 10^{-11}$& $-1.97 \cdot 10^{-11}$\\
    10$^{10}$ & $-1.34 \cdot 10^{-13}$& $-2.01 \cdot 10^{-13}$& $-1.97 \cdot 10^{-13}$\\
    \hline
    \end{tabular}
    \label{tabWeN_baf}
\end{table}

The calculated values of $\WaxeN(m_a)$ for the BaF molecule are given in Table~\ref{tabWeN_baf} using three electronic structure computational models in order of increasing accuracy: Dirac-Hartree-Fock (DHF), relativistic coupled cluster with single and double excitation amplitudes (CCSD), and CCSD(T). Notably, electron correlation effects, i.e., effects beyond the DHF approximation, are significant for the molecular constant $\WaxeN(m_a)$, and their importance increases with the ALP's mass. Comparing the ``CCSD'' and ``CCSD(T)'' columns, one can see good convergence with respect to the treatment of electron correlation effects.
The difference of the CCSD(T) and CCSD values can be used as a measure of the uncertainty in the treatment of electron correlation effects. We also took the half of the difference of results obtained within the AE3Z and AE2Z basis sets at the CCSD(T) level as a measure of the uncertainty due to the basis set incompleteness (see also Ref.~\cite{Maison:2022} 
and Table III of  Supplementary Material for the results in a smaller basis set AE2Z).
The overall uncertainty of the calculated $\WaxeN(m_a)$ values is less than 5\%.

Similar to the YbOH molecule~\cite{Maison:2021} and the HfF$^+$ molecular cation~\cite{Prosnyak:2023a}, the parameter $\WaxeN(m_a)$ remains almost independent of $m_a$ for ALPs with $m_a \leq 10^2$~eV and undergoes a sign change in the interval between $m_a = 10^4$ eV and $m_a = 10^5$ eV.
According to our calculations, the extended nucleus size effect in the BaF molecule depends on $m_a$ similarly to the HfF$^+$ case described above, but smaller by magnitude: starting from the negligible value for ALPs with $m_a < 10^7$ eV, the effect 
appears at $m_a = 10^7$~eV and reaches
a magnitude of about 6\% for $m_a = 10^{10}$~eV and stabilizes 
(see Table IV of the Supplementary Material).

The calculated values of $\Waxee(m_a)$ for the BaF molecule can be found in Table~\ref{tabWee}. Similar to $\WaxeN(m_a)$, electron correlation effects become more pronounced with increasing ALP mass, although they remain significant even for light ALPs. The convergence with respect to the treatment of electron correlation effects, as it can be seen from the ``CCSD'' and ``CCSD(T)'' columns, is good. 
Due to its complexity, we calculated the $\Waxee(m_a)$ molecular parameters within one basis set only. To be conservative, we (arbitrary) assumed that the basis set uncertainty for $\Waxee(m_a)$ may be twice as large as that for $\WaxeN(m_a)$, resulting in an overall uncertainty of approximately 10\% for $\Waxee(m_a)$.

One can see from Table~\ref{tabWee}, that for light ALPs with masses $m_a \leq 10^2$ eV, the parameter $\Waxee(m_a)$ remains nearly constant.
As the ALP mass increases, one can see a noticeable change in $\Waxee(m_a)$ for $m_a=10^3$ eV, followed by a drop of approximately by a factor of 3 for $m_a=10^4$ eV. Finally, in the range between $m_a = 10^5$ eV and $m_a = 10^6$ eV, the parameter $\Waxee(m_a)$ undergoes a change in sign.

\begin{table}[h]
    \caption{The calculated values of $\Waxee(m_a)$ molecular parameters (in units of $m_e c / \hbar$) at various levels of theory. 
    }
    \centering
    \begin{tabular}{lccr}
    \hline
    \hline
    $m_a$, eV  & DHF & CCSD & CCSD(T) \\
    
    \hline
    1  & $+6.3 \cdot 10^{-6}$&  $+8.1 \cdot 10^{-6}$&     $+8.0 \cdot 10^{-6}$\\
    10 & $+6.3 \cdot 10^{-6}$&  $+8.1 \cdot 10^{-6}$&     $+8.0 \cdot 10^{-6}$\\
    $10^2$ & $+6.3 \cdot 10^{-6}$&  $+8.1 \cdot 10^{-6}$& $+8.0 \cdot 10^{-6}$\\
    $10^3$ & $+4.8 \cdot 10^{-6}$&  $+6.6 \cdot 10^{-6}$& $+6.5 \cdot 10^{-6}$\\
    $10^4$ & $+1.2 \cdot 10^{-6}$&  $+2.0 \cdot 10^{-6}$& $+2.0 \cdot 10^{-6}$\\
    $10^5$ & $+9.3 \cdot 10^{-8}$&  $+1.4 \cdot 10^{-7}$& $+1.4 \cdot 10^{-7}$\\
    $10^6$ & $-3.3 \cdot 10^{-9}$&  $-5.2 \cdot 10^{-9}$& $-5.1 \cdot 10^{-9}$\\
    $10^7$ & $-5.1 \cdot 10^{-11}$&  $-8.0 \cdot 10^{-11}$& $-7.8 \cdot 10^{-11}$\\
    $10^8$ & $-5.1 \cdot 10^{-13}$&  $-8.0 \cdot 10^{-13}$& $-7.9 \cdot 10^{-13}$\\
    $10^9$ & $-5.1 \cdot 10^{-15}$&  $-8.0 \cdot 10^{-15}$& $-7.9 \cdot 10^{-15}$\\
    $10^{10}$ & $-5.1 \cdot 10^{-17}$& $-8.1 \cdot 10^{-17}$& $-7.9 \cdot 10^{-17}$\\
    \hline
    \end{tabular}
    \label{tabWee}
\end{table}

\begin{figure}[htp]
\begin{center}
(a)\includegraphics[width=1.0\linewidth]{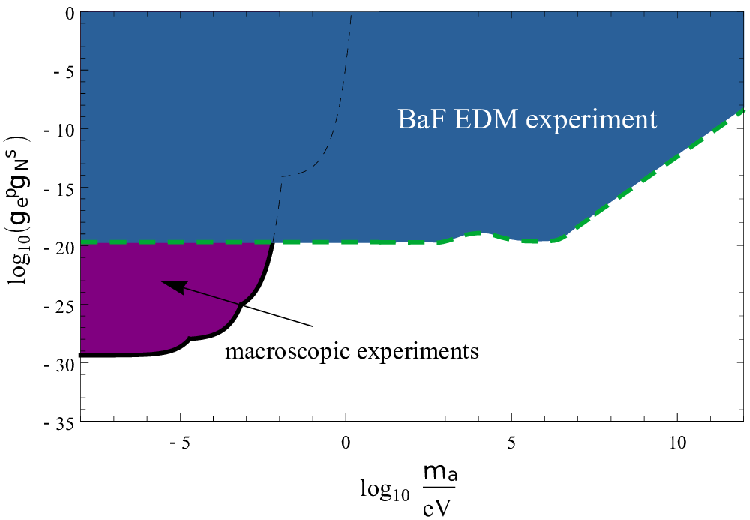}
(b)\includegraphics[width=1.0\linewidth]{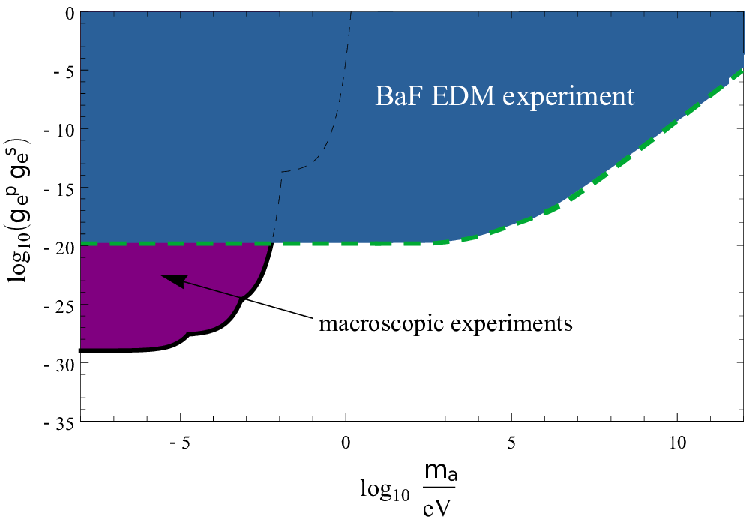}
\end{center}
\caption{
Derived constraints on (a) $|\bar{g}_N^s g_e^p|$ and on (b) $|g_e^s g_e^p|$ products (dashed green line and blue region) corresponding to the $e$EDM limitation $|\de|= 5.0\times 10^{-30}$  $\ensuremath{e \cdot {\rm cm}}$ expected in the planned experiment on BaF~\cite{BaF:2018}. At the graph scale, the constraints almost coincide with the current limitations on $|\bar{g}_N^s g_e^p|$ and $|g_e^s g_e^p|$ obtained in the HfF$^+$ experiment~\cite{newlimit1,Prosnyak:2023a}. Constraint on $|\bar{g}_N^s g_e^p|$ from macroscopic experiments (black line and violet region) were compiled in Ref.~\cite{Stadnik:2018} using the results of Refs.~\cite{youdin1996limits, ni1999search, hammond2007new, hoedl2011improved, terrano2015short, crescini2017improved, rong2018searching}. 
The macroscopic constraints of $|g_e^s g_e^p|$ were estimated in Ref.~\cite{Stadnik:2018} by reevaluating the published constraints on $|\bar{g}_N^s g_e^p|$.
}
\label{figure:1}
\end{figure}

The experiment to measure \Ti,\Par-violating effects on the BaF molecule is under preparation~\cite{BaF:2018}. Using the expected constraint on the magnitude of the $e$EDM $|\de|= 5.0\times 10^{-30} e\cdot$cm~\cite{BaF:2018} and the theoretical value of the effective field~$E_{\rm eff}=W_d|\Omega|\approx 1.565 \cdot 10^{24}$ Hz/(e$\cdot$cm)~\cite{BaF_enhancement_2021}, it is possible to estimate the expected sensitivity of the experiment in terms of the \Ti,\Par-violating energy shift: 
$\delta E \approx 8\text{ } \mu\text{Hz}$.
According to Eqs.~(\ref{Etp_eN}) and~(\ref{Etp_ee}), this constraint can be reinterpreted in terms of limits on the products of ALP coupling constants. The obtained results are given in Table~\ref{tab_limits} and in Fig.~\ref{figure:1}.
Let us consider the two limiting cases separately.

\begin{table}[h]
    \caption{The calculated constraints on $|\bar{g}_N^s g_e^p|$ and $|g_e^s g_e^p|$ products corresponding to the $e$EDM limitation $|\de|= 5.0\times 10^{-30}$  $\ensuremath{e \cdot {\rm cm}}$ expected in the planned experiment on BaF~\cite{BaF:2018}.}
    \centering
    \begin{tabular}{lcc}
    \hline
    \hline
    $m_a$, eV  & $|\bar{g}_N^s g_e^p|$, $\hbar c$& $|g_e^s g_e^p|$, $\hbar c$\\
    
    \hline
    1  & $+7.29\cdot 10^{-21}$&  $+1.6 \cdot 10^{-20}$\\
    10 & $+7.29 \cdot 10^{-21}$&  $+1.6 \cdot 10^{-20}$\\
    $10^2$ & $+7.31 \cdot 10^{-21}$&  $+1.6 \cdot 10^{-20}$\\
    $10^3$ & $+8.71 \cdot 10^{-21}$&  $+1.9 \cdot 10^{-20}$\\
    $10^4$ & $+6.81 \cdot 10^{-21}$&  $+6.5 \cdot 10^{-20}$\\
    $10^5$ & $+1.18 \cdot 10^{-20}$&  $+9.2 \cdot 10^{-19}$\\
    $10^6$ & $+2.69 \cdot 10^{-20}$&  $+2.5 \cdot 10^{-17}$\\
    $10^7$ & $+9.34 \cdot 10^{-19}$&  $+1.6 \cdot 10^{-15}$\\
    $10^8$ & $+6.65 \cdot 10^{-17}$&  $+1.6 \cdot 10^{-13}$\\
    $10^9$ & $+6.43 \cdot 10^{-15}$&  $+1.6 \cdot 10^{-11}$\\
    $10^{10}$ & $+6.43 \cdot 10^{-13}$& $+1.6 \cdot 10^{-9}$\\
    \hline
    \end{tabular}
    \label{tab_limits}
\end{table}

\textit{Low-mass limit.} 
As one can see from Tables ~\ref{tabWeN_baf} and ~\ref{tabWee}, molecular parameters $W_{ax}^{eN}(m_a)$ and $W_{ax}^{ee}(m_a)$ are almost ALP-mass-independent for light ALPs with masses $m_a \leq 10^2$ eV. This can be explained by the fact that, in this case, the range of the interaction is comparable to the size of the molecule. Therefore, in the Yukawa-type interaction one can approximate the exponential factor by unity and the interaction becomes
independent of $m_a$. Thus, for light ALPs, constraints on $|\bar{g}_N^s g_e^p|$ and $|g_e^s g_e^p|$ become mass independent.

\textit{High-mass limit.} For ALPs with high masses ($m_a \geq 10^9$ eV) the interaction becomes very short-range and the following approximation in Eq.~(\ref{WaxeNME}) can be made:
\begin{equation}
\label{eN_delta_potential}
   \int d\boldsymbol{R} \rho(\boldsymbol{R})
    V_{eN}(\boldsymbol{r}_i-\boldsymbol{R})  \approx
 +i \frac{g_N^sg_e^p}{m_a^2} \rho(r_i) \gamma_0 \gamma_5
\end{equation}
and the following approximate relations for $\WaxeN$ and $\Waxee$ can be obtained (see also Refs.~\cite{Stadnik:2018,Maison:2021}): $\WaxeN(m_a) \simeq \widetilde{W}^{eN} m_a^{-2}$ and $\Waxee(m_a) \simeq \widetilde{W}^{ee} m_a^{-2}$, where parameters $\widetilde{W}^{eN}$ and $\widetilde{W}^{ee}$ are independent of $m_a$. Thus, the energy shift for the electron-nuclear interaction can be parameterized as follows:
\begin{equation} 
    \delta E \approx  \frac{\bar{g}_N^s g_e^p}{m_a^2} \Omega\widetilde{W}^{eN},
\end{equation}
where
\begin{equation}
\widetilde{W}^{eN} = \lim_{m_a \rightarrow +\infty} m_a^2 \WaxeN(m_a).
\end{equation}
A similar dependence can be used for the electron-electron interaction: 
\begin{equation} 
    \delta E \approx  \frac{g_e^s g_e^p}{m_a^2} \Omega\widetilde{W}^{ee},
\end{equation}
where
\begin{equation}
\widetilde{W}^{ee} = \lim_{m_a \rightarrow +\infty} m_a^2  \Waxee(m_a).
\end{equation}

In the case of heavy ALPs, the electron-nucleus interaction parameter (\ref{WaxeNME}) can be related to the parameter of the scalar-pseudoscalar nucleus-electron interaction (see also Eq.~(\ref{eN_delta_potential})). Indeed, the potential of the latter interaction can be expressed in the following form~\cite{GFreview}:
\begin{equation}
    V_{S-PS} =  i Z k_{S-PS} \frac{G_F}{\sqrt{2}}\rho(\boldsymbol{r})\gamma_0 \gamma_5,
\end{equation}
where $Z$ is the charge of the nucleus, $k_{S-PS}$ is the dimensionless coupling constant, $G_F$ is the Fermi coupling constant ($2.2225 \cdot 10^{-14}$ in atomic units).  The energy shift corresponding to this interaction is 
\begin{equation}
    \delta E = \Omega k_{S-PS} W_{S-PS},
\end{equation}
where 
\begin{equation}
    W_{S-PS} = 
    \frac{1}{\Omega}
    \langle \Psi | 
    \frac{1}{k_{S-PS}}\sum_{i=1}^{N_e}{V_{S-PS}(\boldsymbol{r}_i)}
    | \Psi \rangle.
\end{equation}
Therefore, the approximate relation between the molecular parameters can be obtained:
\begin{equation}
    \WaxeN(m_a \to \infty) = \frac{A \sqrt{2}}{Z}\frac{1}{G_F m^2_a}W_{S-PS}
\end{equation}
In Ref.~\cite{BaF_enhancement_2021} the value of the molecular parameter 
$W_{S-PS}=8.29$~kHz was calculated. It corresponds to 
$\WaxeN(m_a = 10^{10}\text{eV}) = -2.01 \cdot 10^{-13} m_e c/ \hbar$, 
which is very close to the value $-1.97 \cdot 10^{-13} m_e c/ \hbar$ obtained in the present study (see Table~\ref{tabWeN_baf}).

For convenience, the constraints obtained for both light and heavy ALPs are summarized in Table~\ref{TConstraints}. In the case of light ALPs, the expected limitations on the products $|\bar{g}_{N}^s g_e^p|$ and $|g_e^s g_e^p|$ are approximately 50\% and 40\% better, respectively, than those derived~\cite{Prosnyak:2023a} from the HfF$^+$ experiment~\cite{newlimit1}. The constraint on $|\bar{g}_N^s g_e^p|/m_a^2$ is almost twice weaker, while the constraint on $|g_e^s g_e^p|/m_a^2$ is about 5\% better for heavy ALPs compared to the HfF$^+$ case~\cite{Prosnyak:2023a}.
Note that these comparisons are based on the present limitations of the \Ti,\Par-violating energy shift in HfF$^+$~\cite{newlimit1}, corresponding to $|\de| < 4.1 \times 10^{-30}$ $\ensuremath{e \cdot {\rm cm}}$, and the expected sensitivity of the BaF experiment~\cite{BaF:2018}, corresponding to $|\de|= 5.0\times 10^{-30}$ $\ensuremath{e \cdot {\rm cm}}$. Taking into account the very close limitations on the $e$EDM, the numbers given above correspond to the relative sensitivities of HfF$^+$ and BaF molecules to different \Ti,\Par-violating effects. The difference in sensitivities is important to set more robust constraints on various sources of \Ti,\Par-violating effects~\cite{Jung:13}.
As one can see, the expected limitations on the products of the considered ALP-coupling constants from the experiment on the BaF molecule are of the same order of magnitude as those from the experiment on the HfF${}^+$ molecular cation. Thus, the conclusions drawn from the comparison with macroscopic experiments (see Ref.~\cite{Hare:2020} for a review of such experiments) will be nearly identical to those for the HfF$^+$ case~\cite{Prosnyak:2023a}: for very light ALPs, much better constraints can be obtained from macroscopic experiments~\cite{Hare:2020,Heckel:2008,Wineland:1991,Lee:2018,hoedl2011improved,XENON1T:2019,youdin1996limits, ni1999search, hammond2007new, terrano2015short, crescini2017improved, rong2018searching}. However, for heavy ALPs ($m_a > 10^{-3}$ eV), the limitations that can be achieved from the planned experiment on the BaF molecule~\cite{BaF:2018} can be more stringent, see Fig.~\ref{figure:1}, which updates the plot given in Ref.~\cite{Stadnik:2018}. 

\begin{table}[!h]
\caption{Summary of the obtained constraints on the combinations of ALP coupling constants for limiting cases of light- and high-mass ALPs. These constraints correspond to the expected sensitivity of the planned experiment on BaF~\cite{BaF:2018}.}
\centering
\begin{tabular}{ll}
\hline
\hline
Limit                                                        & Value                 \\
\hline
$|\bar{g}_{N}^s g_e^p|/(\hbar c)$, $m_a \ll 1\ \textrm{keV}$     & $7.29 \times 10^{-21}$\\
$|g_e^s g_e^p|/(\hbar c)$, $m_a \ll 1\ \textrm{keV}$            & $1.6 \times 10^{-20}$\\
$|\bar{g}_N^s g_e^p|/(\hbar c m_a^{2})$, $m_a \ge 1\ \textrm{GeV}$  & $6.43 \cdot 10^{-15}~\textrm{GeV}^{-2}$\\
$|g_e^s g_e^p|/(\hbar c m_a^{2})$, $m_a \ge 1\ \textrm{GeV}$        & $1.6 \cdot 10^{-11}~\textrm{GeV}^{-2}$\\
\hline
\hline
\end{tabular}
\label{TConstraints}
\end{table}

\section{Conclusions}

The magnitude of the effects of the axion-like-particles-mediated scalar-pseudoscalar \Ti,\Par-violating nucleus-electron and electron-electron interactions were estimated for the barium monofluoride molecule in the $X^2\Sigma_{1/2}$ ground electronic state. 
We demonstrated that the extended nucleus size effect is valuable in case of the nucleus-electron interaction mediated by heavy ALPs and should be taken into account in accurate treatment.

In our calculations, we obtained the molecular parameters that characterize the \Ti,\Par-violating interactions in the ground state of the BaF molecule. These parameters can be used for interpretation of such experiment in terms of the scalar-pseudoscalar ALP-mediated interactions with arbitrary mass ALPs of future experiment~\cite{BaF:2018}. The developed approach can be used in future studies of the scalar-pseudoscalar ALP-mediated nucleus-electron interaction in other molecules and atoms.

\section{Acknowledgments}
The authors are grateful to D. E. Maison for valuable  discussions.
Electronic structure calculations were carried out using computing resources of the federal collective usage center Complex for Simulation and Data Processing for Mega-science Facilities at National Research Centre ``Kurchatov Institute'', http://ckp.nrcki.ru/, and partly using the computing resources of the quantum chemistry laboratory.

Molecular coupled cluster electronic structure calculations were supported by the Russian Science Foundation Grant No. 19-72-10019-P (https://rscf.ru/project/22-72-41010/). Calculations of the $\WaxeN(m_a)$ matrix elements and development of corresponding code were supported by the Foundation for the Advancement of Theoretical Physics and Mathematics ``BASIS'' Grant according to Projects No. 21-1-2-47-1.


\begin{thebibliography}{65}%
\makeatletter
\providecommand \@ifxundefined [1]{%
 \@ifx{#1\undefined}
}%
\providecommand \@ifnum [1]{%
 \ifnum #1\expandafter \@firstoftwo
 \else \expandafter \@secondoftwo
 \fi
}%
\providecommand \@ifx [1]{%
 \ifx #1\expandafter \@firstoftwo
 \else \expandafter \@secondoftwo
 \fi
}%
\providecommand \natexlab [1]{#1}%
\providecommand \enquote  [1]{``#1''}%
\providecommand \bibnamefont  [1]{#1}%
\providecommand \bibfnamefont [1]{#1}%
\providecommand \citenamefont [1]{#1}%
\providecommand \href@noop [0]{\@secondoftwo}%
\providecommand \href [0]{\begingroup \@sanitize@url \@href}%
\providecommand \@href[1]{\@@startlink{#1}\@@href}%
\providecommand \@@href[1]{\endgroup#1\@@endlink}%
\providecommand \@sanitize@url [0]{\catcode `\\12\catcode `\$12\catcode
  `\&12\catcode `\#12\catcode `\^12\catcode `\_12\catcode `\%12\relax}%
\providecommand \@@startlink[1]{}%
\providecommand \@@endlink[0]{}%
\providecommand \url  [0]{\begingroup\@sanitize@url \@url }%
\providecommand \@url [1]{\endgroup\@href {#1}{\urlprefix }}%
\providecommand \urlprefix  [0]{URL }%
\providecommand \Eprint [0]{\href }%
\providecommand \doibase [0]{http://dx.doi.org/}%
\providecommand \selectlanguage [0]{\@gobble}%
\providecommand \bibinfo  [0]{\@secondoftwo}%
\providecommand \bibfield  [0]{\@secondoftwo}%
\providecommand \translation [1]{[#1]}%
\providecommand \BibitemOpen [0]{}%
\providecommand \bibitemStop [0]{}%
\providecommand \bibitemNoStop [0]{.\EOS\space}%
\providecommand \EOS [0]{\spacefactor3000\relax}%
\providecommand \BibitemShut  [1]{\csname bibitem#1\endcsname}%
\let\auto@bib@innerbib\@empty
\bibitem [{\citenamefont {Safronova}\ \emph {et~al.}(2018)\citenamefont
  {Safronova}, \citenamefont {Budker}, \citenamefont {DeMille}, \citenamefont
  {Kimball}, \citenamefont {Derevianko},\ and\ \citenamefont
  {Clark}}]{Safronova:18}%
  \BibitemOpen
  \bibfield  {author} {\bibinfo {author} {\bibfnamefont {M.~S.}\ \bibnamefont
  {Safronova}}, \bibinfo {author} {\bibfnamefont {D.}~\bibnamefont {Budker}},
  \bibinfo {author} {\bibfnamefont {D.}~\bibnamefont {DeMille}}, \bibinfo
  {author} {\bibfnamefont {D.~F.~J.}\ \bibnamefont {Kimball}}, \bibinfo
  {author} {\bibfnamefont {A.}~\bibnamefont {Derevianko}}, \ and\ \bibinfo
  {author} {\bibfnamefont {C.~W.}\ \bibnamefont {Clark}},\ }\href {\doibase
  10.1103/RevModPhys.90.025008} {\bibfield  {journal} {\bibinfo  {journal}
  {Rev.\ Mod.\ Phys.}\ }\textbf {\bibinfo {volume} {90}},\ \bibinfo {pages}
  {025008} (\bibinfo {year} {2018})}\BibitemShut {NoStop}%
\bibitem [{\citenamefont {Alarcon}\ \emph {et~al.}()\citenamefont {Alarcon},
  \citenamefont {Alexander}, \citenamefont {Anastassopoulos}, \citenamefont
  {Aoki}, \citenamefont {Baartman}, \citenamefont {Baeßler}, \citenamefont
  {Bartoszek}, \citenamefont {Beck}, \citenamefont {Bedeschi}, \citenamefont
  {Berger}, \citenamefont {Berz}, \citenamefont {Bethlem}, \citenamefont
  {Bhattacharya}, \citenamefont {Blaskiewicz}, \citenamefont {Blum},
  \citenamefont {Bowcock}, \citenamefont {Borschevsky}, \citenamefont {Brown},
  \citenamefont {Budker}, \citenamefont {Burdin}, \citenamefont {Casey},
  \citenamefont {Casse}, \citenamefont {Cantatore}, \citenamefont {Cheng},
  \citenamefont {Chupp}, \citenamefont {Cianciolo}, \citenamefont {Cirigliano},
  \citenamefont {Clayton}, \citenamefont {Crawford}, \citenamefont {Das},
  \citenamefont {Davoudiasl}, \citenamefont {de~Vries}, \citenamefont
  {DeMille}, \citenamefont {Denisov}, \citenamefont {Diwan}, \citenamefont
  {Doyle}, \citenamefont {Engel}, \citenamefont {Fanourakis}, \citenamefont
  {Fatemi}, \citenamefont {Filippone}, \citenamefont {Flambaum}, \citenamefont
  {Fleig}, \citenamefont {Fomin}, \citenamefont {Fischer}, \citenamefont
  {Gabrielse}, \citenamefont {Ruiz}, \citenamefont {Gardikiotis}, \citenamefont
  {Gatti}, \citenamefont {Geraci}, \citenamefont {Gooding}, \citenamefont
  {Golub}, \citenamefont {Graham}, \citenamefont {Gray}, \citenamefont
  {Griffith}, \citenamefont {Haciomeroglu}, \citenamefont {Gwinner},
  \citenamefont {Hoekstra}, \citenamefont {Hoffstaetter}, \citenamefont
  {Huang}, \citenamefont {Hutzler}, \citenamefont {Incagli}, \citenamefont
  {Ito}, \citenamefont {Izubuchi}, \citenamefont {Jayich}, \citenamefont
  {Jeong}, \citenamefont {Kaplan}, \citenamefont {Karuza}, \citenamefont
  {Kawall}, \citenamefont {Kim}, \citenamefont {Koop}, \citenamefont {Korsch},
  \citenamefont {Korobkina}, \citenamefont {Lebedev}, \citenamefont {Lee},
  \citenamefont {Lee}, \citenamefont {Lehnert}, \citenamefont {Leung},
  \citenamefont {Liu}, \citenamefont {Long}, \citenamefont {Lusiani},
  \citenamefont {Marciano}, \citenamefont {Maroudas}, \citenamefont
  {Matlashov}, \citenamefont {Matsumoto}, \citenamefont {Mawhorter},
  \citenamefont {Meot}, \citenamefont {Mereghetti}, \citenamefont {Miller},
  \citenamefont {Morse}, \citenamefont {Mott}, \citenamefont {Omarov},
  \citenamefont {Orozco}, \citenamefont {O'Shaughnessy}, \citenamefont {Ozben},
  \citenamefont {Park}, \citenamefont {Pattie}, \citenamefont {Petrov},
  \citenamefont {Piacentino}, \citenamefont {Plaster}, \citenamefont
  {Podobedov}, \citenamefont {Poelker}, \citenamefont {Pocanic}, \citenamefont
  {Prasannaa}, \citenamefont {Price}, \citenamefont {Ramsey-Musolf},
  \citenamefont {Raparia}, \citenamefont {Rajendran}, \citenamefont {Reece},
  \citenamefont {Reid}, \citenamefont {Rescia}, \citenamefont {Ritz},
  \citenamefont {Roberts}, \citenamefont {Safronova}, \citenamefont {Sakemi},
  \citenamefont {Schmidt-Wellenburg}, \citenamefont {Shindler}, \citenamefont
  {Semertzidis}, \citenamefont {Silenko}, \citenamefont {Singh}, \citenamefont
  {Skripnikov}, \citenamefont {Soni}, \citenamefont {Stephenson}, \citenamefont
  {Suleiman}, \citenamefont {Sunaga}, \citenamefont {Syphers}, \citenamefont
  {Syritsyn}, \citenamefont {Tarbutt}, \citenamefont {Thoerngren},
  \citenamefont {Timmermans}, \citenamefont {Tishchenko}, \citenamefont
  {Titov}, \citenamefont {Tsoupas}, \citenamefont {Tzamarias}, \citenamefont
  {Variola}, \citenamefont {Venanzoni}, \citenamefont {Vilella}, \citenamefont
  {Vossebeld}, \citenamefont {Winter}, \citenamefont {Won}, \citenamefont
  {Zelenski}, \citenamefont {Zelevinsky}, \citenamefont {Zhou},\ and\
  \citenamefont {Zioutas}}]{eEDM_snowmass:2022}%
  \BibitemOpen
  \bibfield  {author} {\bibinfo {author} {\bibfnamefont {R.}~\bibnamefont
  {Alarcon}}, \bibinfo {author} {\bibfnamefont {J.}~\bibnamefont {Alexander}},
  \bibinfo {author} {\bibfnamefont {V.}~\bibnamefont {Anastassopoulos}},
  \bibinfo {author} {\bibfnamefont {T.}~\bibnamefont {Aoki}}, \bibinfo {author}
  {\bibfnamefont {R.}~\bibnamefont {Baartman}}, \bibinfo {author}
  {\bibfnamefont {S.}~\bibnamefont {Baeßler}}, \bibinfo {author}
  {\bibfnamefont {L.}~\bibnamefont {Bartoszek}}, \bibinfo {author}
  {\bibfnamefont {D.~H.}\ \bibnamefont {Beck}}, \bibinfo {author}
  {\bibfnamefont {F.}~\bibnamefont {Bedeschi}}, \bibinfo {author}
  {\bibfnamefont {R.}~\bibnamefont {Berger}}, \bibinfo {author} {\bibfnamefont
  {M.}~\bibnamefont {Berz}}, \bibinfo {author} {\bibfnamefont {H.~L.}\
  \bibnamefont {Bethlem}}, \bibinfo {author} {\bibfnamefont {T.}~\bibnamefont
  {Bhattacharya}}, \bibinfo {author} {\bibfnamefont {M.}~\bibnamefont
  {Blaskiewicz}}, \bibinfo {author} {\bibfnamefont {T.}~\bibnamefont {Blum}},
  \bibinfo {author} {\bibfnamefont {T.}~\bibnamefont {Bowcock}}, \bibinfo
  {author} {\bibfnamefont {A.}~\bibnamefont {Borschevsky}}, \bibinfo {author}
  {\bibfnamefont {K.}~\bibnamefont {Brown}}, \bibinfo {author} {\bibfnamefont
  {D.}~\bibnamefont {Budker}}, \bibinfo {author} {\bibfnamefont
  {S.}~\bibnamefont {Burdin}}, \bibinfo {author} {\bibfnamefont {B.~C.}\
  \bibnamefont {Casey}}, \bibinfo {author} {\bibfnamefont {G.}~\bibnamefont
  {Casse}}, \bibinfo {author} {\bibfnamefont {G.}~\bibnamefont {Cantatore}},
  \bibinfo {author} {\bibfnamefont {L.}~\bibnamefont {Cheng}}, \bibinfo
  {author} {\bibfnamefont {T.}~\bibnamefont {Chupp}}, \bibinfo {author}
  {\bibfnamefont {V.}~\bibnamefont {Cianciolo}}, \bibinfo {author}
  {\bibfnamefont {V.}~\bibnamefont {Cirigliano}}, \bibinfo {author}
  {\bibfnamefont {S.~M.}\ \bibnamefont {Clayton}}, \bibinfo {author}
  {\bibfnamefont {C.}~\bibnamefont {Crawford}}, \bibinfo {author}
  {\bibfnamefont {B.~P.}\ \bibnamefont {Das}}, \bibinfo {author} {\bibfnamefont
  {H.}~\bibnamefont {Davoudiasl}}, \bibinfo {author} {\bibfnamefont
  {J.}~\bibnamefont {de~Vries}}, \bibinfo {author} {\bibfnamefont
  {D.}~\bibnamefont {DeMille}}, \bibinfo {author} {\bibfnamefont
  {D.}~\bibnamefont {Denisov}}, \bibinfo {author} {\bibfnamefont {M.~V.}\
  \bibnamefont {Diwan}}, \bibinfo {author} {\bibfnamefont {J.~M.}\ \bibnamefont
  {Doyle}}, \bibinfo {author} {\bibfnamefont {J.}~\bibnamefont {Engel}},
  \bibinfo {author} {\bibfnamefont {G.}~\bibnamefont {Fanourakis}}, \bibinfo
  {author} {\bibfnamefont {R.}~\bibnamefont {Fatemi}}, \bibinfo {author}
  {\bibfnamefont {B.~W.}\ \bibnamefont {Filippone}}, \bibinfo {author}
  {\bibfnamefont {V.~V.}\ \bibnamefont {Flambaum}}, \bibinfo {author}
  {\bibfnamefont {T.}~\bibnamefont {Fleig}}, \bibinfo {author} {\bibfnamefont
  {N.}~\bibnamefont {Fomin}}, \bibinfo {author} {\bibfnamefont
  {W.}~\bibnamefont {Fischer}}, \bibinfo {author} {\bibfnamefont
  {G.}~\bibnamefont {Gabrielse}}, \bibinfo {author} {\bibfnamefont {R.~F.~G.}\
  \bibnamefont {Ruiz}}, \bibinfo {author} {\bibfnamefont {A.}~\bibnamefont
  {Gardikiotis}}, \bibinfo {author} {\bibfnamefont {C.}~\bibnamefont {Gatti}},
  \bibinfo {author} {\bibfnamefont {A.}~\bibnamefont {Geraci}}, \bibinfo
  {author} {\bibfnamefont {J.}~\bibnamefont {Gooding}}, \bibinfo {author}
  {\bibfnamefont {B.}~\bibnamefont {Golub}}, \bibinfo {author} {\bibfnamefont
  {P.}~\bibnamefont {Graham}}, \bibinfo {author} {\bibfnamefont
  {F.}~\bibnamefont {Gray}}, \bibinfo {author} {\bibfnamefont {W.~C.}\
  \bibnamefont {Griffith}}, \bibinfo {author} {\bibfnamefont {S.}~\bibnamefont
  {Haciomeroglu}}, \bibinfo {author} {\bibfnamefont {G.}~\bibnamefont
  {Gwinner}}, \bibinfo {author} {\bibfnamefont {S.}~\bibnamefont {Hoekstra}},
  \bibinfo {author} {\bibfnamefont {G.~H.}\ \bibnamefont {Hoffstaetter}},
  \bibinfo {author} {\bibfnamefont {H.}~\bibnamefont {Huang}}, \bibinfo
  {author} {\bibfnamefont {N.~R.}\ \bibnamefont {Hutzler}}, \bibinfo {author}
  {\bibfnamefont {M.}~\bibnamefont {Incagli}}, \bibinfo {author} {\bibfnamefont
  {T.~M.}\ \bibnamefont {Ito}}, \bibinfo {author} {\bibfnamefont
  {T.}~\bibnamefont {Izubuchi}}, \bibinfo {author} {\bibfnamefont {A.~M.}\
  \bibnamefont {Jayich}}, \bibinfo {author} {\bibfnamefont {H.}~\bibnamefont
  {Jeong}}, \bibinfo {author} {\bibfnamefont {D.}~\bibnamefont {Kaplan}},
  \bibinfo {author} {\bibfnamefont {M.}~\bibnamefont {Karuza}}, \bibinfo
  {author} {\bibfnamefont {D.}~\bibnamefont {Kawall}}, \bibinfo {author}
  {\bibfnamefont {O.}~\bibnamefont {Kim}}, \bibinfo {author} {\bibfnamefont
  {I.}~\bibnamefont {Koop}}, \bibinfo {author} {\bibfnamefont {W.}~\bibnamefont
  {Korsch}}, \bibinfo {author} {\bibfnamefont {E.}~\bibnamefont {Korobkina}},
  \bibinfo {author} {\bibfnamefont {V.}~\bibnamefont {Lebedev}}, \bibinfo
  {author} {\bibfnamefont {J.}~\bibnamefont {Lee}}, \bibinfo {author}
  {\bibfnamefont {S.}~\bibnamefont {Lee}}, \bibinfo {author} {\bibfnamefont
  {R.}~\bibnamefont {Lehnert}}, \bibinfo {author} {\bibfnamefont {K.~K.~H.}\
  \bibnamefont {Leung}}, \bibinfo {author} {\bibfnamefont {C.-Y.}\ \bibnamefont
  {Liu}}, \bibinfo {author} {\bibfnamefont {J.}~\bibnamefont {Long}}, \bibinfo
  {author} {\bibfnamefont {A.}~\bibnamefont {Lusiani}}, \bibinfo {author}
  {\bibfnamefont {W.~J.}\ \bibnamefont {Marciano}}, \bibinfo {author}
  {\bibfnamefont {M.}~\bibnamefont {Maroudas}}, \bibinfo {author}
  {\bibfnamefont {A.}~\bibnamefont {Matlashov}}, \bibinfo {author}
  {\bibfnamefont {N.}~\bibnamefont {Matsumoto}}, \bibinfo {author}
  {\bibfnamefont {R.}~\bibnamefont {Mawhorter}}, \bibinfo {author}
  {\bibfnamefont {F.}~\bibnamefont {Meot}}, \bibinfo {author} {\bibfnamefont
  {E.}~\bibnamefont {Mereghetti}}, \bibinfo {author} {\bibfnamefont {J.~P.}\
  \bibnamefont {Miller}}, \bibinfo {author} {\bibfnamefont {W.~M.}\
  \bibnamefont {Morse}}, \bibinfo {author} {\bibfnamefont {J.}~\bibnamefont
  {Mott}}, \bibinfo {author} {\bibfnamefont {Z.}~\bibnamefont {Omarov}},
  \bibinfo {author} {\bibfnamefont {L.~A.}\ \bibnamefont {Orozco}}, \bibinfo
  {author} {\bibfnamefont {C.~M.}\ \bibnamefont {O'Shaughnessy}}, \bibinfo
  {author} {\bibfnamefont {C.}~\bibnamefont {Ozben}}, \bibinfo {author}
  {\bibfnamefont {S.}~\bibnamefont {Park}}, \bibinfo {author} {\bibfnamefont
  {R.~W.}\ \bibnamefont {Pattie}}, \bibinfo {author} {\bibfnamefont {A.~N.}\
  \bibnamefont {Petrov}}, \bibinfo {author} {\bibfnamefont {G.~M.}\
  \bibnamefont {Piacentino}}, \bibinfo {author} {\bibfnamefont {B.~R.}\
  \bibnamefont {Plaster}}, \bibinfo {author} {\bibfnamefont {B.}~\bibnamefont
  {Podobedov}}, \bibinfo {author} {\bibfnamefont {M.}~\bibnamefont {Poelker}},
  \bibinfo {author} {\bibfnamefont {D.}~\bibnamefont {Pocanic}}, \bibinfo
  {author} {\bibfnamefont {V.~S.}\ \bibnamefont {Prasannaa}}, \bibinfo {author}
  {\bibfnamefont {J.}~\bibnamefont {Price}}, \bibinfo {author} {\bibfnamefont
  {M.~J.}\ \bibnamefont {Ramsey-Musolf}}, \bibinfo {author} {\bibfnamefont
  {D.}~\bibnamefont {Raparia}}, \bibinfo {author} {\bibfnamefont
  {S.}~\bibnamefont {Rajendran}}, \bibinfo {author} {\bibfnamefont
  {M.}~\bibnamefont {Reece}}, \bibinfo {author} {\bibfnamefont
  {A.}~\bibnamefont {Reid}}, \bibinfo {author} {\bibfnamefont {S.}~\bibnamefont
  {Rescia}}, \bibinfo {author} {\bibfnamefont {A.}~\bibnamefont {Ritz}},
  \bibinfo {author} {\bibfnamefont {B.~L.}\ \bibnamefont {Roberts}}, \bibinfo
  {author} {\bibfnamefont {M.~S.}\ \bibnamefont {Safronova}}, \bibinfo {author}
  {\bibfnamefont {Y.}~\bibnamefont {Sakemi}}, \bibinfo {author} {\bibfnamefont
  {P.}~\bibnamefont {Schmidt-Wellenburg}}, \bibinfo {author} {\bibfnamefont
  {A.}~\bibnamefont {Shindler}}, \bibinfo {author} {\bibfnamefont {Y.~K.}\
  \bibnamefont {Semertzidis}}, \bibinfo {author} {\bibfnamefont
  {A.}~\bibnamefont {Silenko}}, \bibinfo {author} {\bibfnamefont {J.~T.}\
  \bibnamefont {Singh}}, \bibinfo {author} {\bibfnamefont {L.~V.}\ \bibnamefont
  {Skripnikov}}, \bibinfo {author} {\bibfnamefont {A.}~\bibnamefont {Soni}},
  \bibinfo {author} {\bibfnamefont {E.}~\bibnamefont {Stephenson}}, \bibinfo
  {author} {\bibfnamefont {R.}~\bibnamefont {Suleiman}}, \bibinfo {author}
  {\bibfnamefont {A.}~\bibnamefont {Sunaga}}, \bibinfo {author} {\bibfnamefont
  {M.}~\bibnamefont {Syphers}}, \bibinfo {author} {\bibfnamefont
  {S.}~\bibnamefont {Syritsyn}}, \bibinfo {author} {\bibfnamefont {M.~R.}\
  \bibnamefont {Tarbutt}}, \bibinfo {author} {\bibfnamefont {P.}~\bibnamefont
  {Thoerngren}}, \bibinfo {author} {\bibfnamefont {R.~G.~E.}\ \bibnamefont
  {Timmermans}}, \bibinfo {author} {\bibfnamefont {V.}~\bibnamefont
  {Tishchenko}}, \bibinfo {author} {\bibfnamefont {A.~V.}\ \bibnamefont
  {Titov}}, \bibinfo {author} {\bibfnamefont {N.}~\bibnamefont {Tsoupas}},
  \bibinfo {author} {\bibfnamefont {S.}~\bibnamefont {Tzamarias}}, \bibinfo
  {author} {\bibfnamefont {A.}~\bibnamefont {Variola}}, \bibinfo {author}
  {\bibfnamefont {G.}~\bibnamefont {Venanzoni}}, \bibinfo {author}
  {\bibfnamefont {E.}~\bibnamefont {Vilella}}, \bibinfo {author} {\bibfnamefont
  {J.}~\bibnamefont {Vossebeld}}, \bibinfo {author} {\bibfnamefont
  {P.}~\bibnamefont {Winter}}, \bibinfo {author} {\bibfnamefont
  {E.}~\bibnamefont {Won}}, \bibinfo {author} {\bibfnamefont {A.}~\bibnamefont
  {Zelenski}}, \bibinfo {author} {\bibfnamefont {T.}~\bibnamefont
  {Zelevinsky}}, \bibinfo {author} {\bibfnamefont {Y.}~\bibnamefont {Zhou}}, \
  and\ \bibinfo {author} {\bibfnamefont {K.}~\bibnamefont {Zioutas}},\ }\href
  {\doibase 10.48550/ARXIV.2203.08103} {\enquote {\bibinfo {title}
  {arxiv:2203.08103 [hep-ph] (2022)},}\ }\BibitemShut {NoStop}%
\bibitem [{\citenamefont {Safronova}(2023)}]{Safronova2023}%
  \BibitemOpen
  \bibfield  {author} {\bibinfo {author} {\bibfnamefont {M.~S.}\ \bibnamefont
  {Safronova}},\ }\enquote {\bibinfo {title} {Searches for new physics},}\ in\
  \href {\doibase 10.1007/978-3-030-73893-8_32} {\emph {\bibinfo {booktitle}
  {Springer Handbook of Atomic, Molecular, and Optical Physics}}},\ \bibinfo
  {editor} {edited by\ \bibinfo {editor} {\bibfnamefont {G.~W.~F.}\
  \bibnamefont {Drake}}}\ (\bibinfo  {publisher} {Springer International
  Publishing},\ \bibinfo {address} {Cham},\ \bibinfo {year} {2023})\ pp.\
  \bibinfo {pages} {471--484}\BibitemShut {NoStop}%
\bibitem [{\citenamefont {Ema}\ \emph {et~al.}(2022)\citenamefont {Ema},
  \citenamefont {Gao},\ and\ \citenamefont {Pospelov}}]{eEDMLimit:2022}%
  \BibitemOpen
  \bibfield  {author} {\bibinfo {author} {\bibfnamefont {Y.}~\bibnamefont
  {Ema}}, \bibinfo {author} {\bibfnamefont {T.}~\bibnamefont {Gao}}, \ and\
  \bibinfo {author} {\bibfnamefont {M.}~\bibnamefont {Pospelov}},\ }\href
  {\doibase 10.1103/PhysRevLett.129.231801} {\bibfield  {journal} {\bibinfo
  {journal} {Phys. Rev. Lett.}\ }\textbf {\bibinfo {volume} {129}},\ \bibinfo
  {pages} {231801} (\bibinfo {year} {2022})}\BibitemShut {NoStop}%
\bibitem [{\citenamefont {Yamaguchi}\ and\ \citenamefont
  {Yamanaka}(2021)}]{Yamaguchi:2021}%
  \BibitemOpen
  \bibfield  {author} {\bibinfo {author} {\bibfnamefont {Y.}~\bibnamefont
  {Yamaguchi}}\ and\ \bibinfo {author} {\bibfnamefont {N.}~\bibnamefont
  {Yamanaka}},\ }\href {\doibase 10.1103/PhysRevD.103.013001} {\bibfield
  {journal} {\bibinfo  {journal} {Phys. Rev. D}\ }\textbf {\bibinfo {volume}
  {103}},\ \bibinfo {pages} {013001} (\bibinfo {year} {2021})}\BibitemShut
  {NoStop}%
\bibitem [{\citenamefont {Engel}\ \emph {et~al.}(2013)\citenamefont {Engel},
  \citenamefont {Ramsey-Musolf},\ and\ \citenamefont {van Kolck}}]{Engel:2013}%
  \BibitemOpen
  \bibfield  {author} {\bibinfo {author} {\bibfnamefont {J.}~\bibnamefont
  {Engel}}, \bibinfo {author} {\bibfnamefont {M.~J.}\ \bibnamefont
  {Ramsey-Musolf}}, \ and\ \bibinfo {author} {\bibfnamefont {U.}~\bibnamefont
  {van Kolck}},\ }\href {\doibase https://doi.org/10.1016/j.ppnp.2013.03.003}
  {\bibfield  {journal} {\bibinfo  {journal} {Progr. Part. Nuc. Phys.}\
  }\textbf {\bibinfo {volume} {71}},\ \bibinfo {pages} {21 } (\bibinfo {year}
  {2013})}\BibitemShut {NoStop}%
\bibitem [{\citenamefont {Chubukov}\ and\ \citenamefont
  {Labzowsky}(2016)}]{Chubukov:2016}%
  \BibitemOpen
  \bibfield  {author} {\bibinfo {author} {\bibfnamefont {D.~V.}\ \bibnamefont
  {Chubukov}}\ and\ \bibinfo {author} {\bibfnamefont {L.~N.}\ \bibnamefont
  {Labzowsky}},\ }\href {\doibase 10.1103/PhysRevA.93.062503} {\bibfield
  {journal} {\bibinfo  {journal} {Phys. Rev. A}\ }\textbf {\bibinfo {volume}
  {93}},\ \bibinfo {pages} {062503} (\bibinfo {year} {2016})}\BibitemShut
  {NoStop}%
\bibitem [{\citenamefont {Commins}(1998)}]{Commins:98}%
  \BibitemOpen
  \bibfield  {author} {\bibinfo {author} {\bibfnamefont {E.~D.}\ \bibnamefont
  {Commins}},\ }\href {\doibase 10.1016/S1049-250X(08)60110-X} {\bibfield
  {journal} {\bibinfo  {journal} {Adv.\ At.\ Mol.\ Opt.\ Phys.}\ }\textbf
  {\bibinfo {volume} {40}},\ \bibinfo {pages} {1} (\bibinfo {year}
  {1998})}\BibitemShut {NoStop}%
\bibitem [{\citenamefont {Chupp}\ \emph {et~al.}(2019)\citenamefont {Chupp},
  \citenamefont {Fierlinger}, \citenamefont {Ramsey-Musolf},\ and\
  \citenamefont {Singh}}]{Chupp:2019}%
  \BibitemOpen
  \bibfield  {author} {\bibinfo {author} {\bibfnamefont {T.~E.}\ \bibnamefont
  {Chupp}}, \bibinfo {author} {\bibfnamefont {P.}~\bibnamefont {Fierlinger}},
  \bibinfo {author} {\bibfnamefont {M.~J.}\ \bibnamefont {Ramsey-Musolf}}, \
  and\ \bibinfo {author} {\bibfnamefont {J.~T.}\ \bibnamefont {Singh}},\ }\href
  {\doibase 10.1103/RevModPhys.91.015001} {\bibfield  {journal} {\bibinfo
  {journal} {Rev. Mod. Phys.}\ }\textbf {\bibinfo {volume} {91}},\ \bibinfo
  {pages} {015001} (\bibinfo {year} {2019})}\BibitemShut {NoStop}%
\bibitem [{\citenamefont {Roussy}\ \emph {et~al.}(2023)\citenamefont {Roussy},
  \citenamefont {Caldwell}, \citenamefont {Wright}, \citenamefont {Cairncross},
  \citenamefont {Shagam}, \citenamefont {Ng}, \citenamefont {Schlossberger},
  \citenamefont {Park}, \citenamefont {Wang}, \citenamefont {Ye},\ and\
  \citenamefont {Cornell}}]{newlimit1}%
  \BibitemOpen
  \bibfield  {author} {\bibinfo {author} {\bibfnamefont {T.~S.}\ \bibnamefont
  {Roussy}}, \bibinfo {author} {\bibfnamefont {L.}~\bibnamefont {Caldwell}},
  \bibinfo {author} {\bibfnamefont {T.}~\bibnamefont {Wright}}, \bibinfo
  {author} {\bibfnamefont {W.~B.}\ \bibnamefont {Cairncross}}, \bibinfo
  {author} {\bibfnamefont {Y.}~\bibnamefont {Shagam}}, \bibinfo {author}
  {\bibfnamefont {K.~B.}\ \bibnamefont {Ng}}, \bibinfo {author} {\bibfnamefont
  {N.}~\bibnamefont {Schlossberger}}, \bibinfo {author} {\bibfnamefont {S.~Y.}\
  \bibnamefont {Park}}, \bibinfo {author} {\bibfnamefont {A.}~\bibnamefont
  {Wang}}, \bibinfo {author} {\bibfnamefont {J.}~\bibnamefont {Ye}}, \ and\
  \bibinfo {author} {\bibfnamefont {E.~A.}\ \bibnamefont {Cornell}},\ }\href
  {\doibase 10.1126/science.adg4084} {\bibfield  {journal} {\bibinfo  {journal}
  {Science}\ }\textbf {\bibinfo {volume} {381}},\ \bibinfo {pages} {46}
  (\bibinfo {year} {2023})}\BibitemShut {NoStop}%
\bibitem [{\citenamefont {Andreev}\ \emph {et~al.}(2018)\citenamefont
  {Andreev}, \citenamefont {Ang}, \citenamefont {DeMille}, \citenamefont
  {Doyle}, \citenamefont {Gabrielse}, \citenamefont {Haefner}, \citenamefont
  {Hutzler}, \citenamefont {Lasner}, \citenamefont {Meisenhelder},
  \citenamefont {O'Leary} \emph {et~al.}}]{ACME:18}%
  \BibitemOpen
  \bibfield  {author} {\bibinfo {author} {\bibfnamefont {V.}~\bibnamefont
  {Andreev}}, \bibinfo {author} {\bibfnamefont {D.}~\bibnamefont {Ang}},
  \bibinfo {author} {\bibfnamefont {D.}~\bibnamefont {DeMille}}, \bibinfo
  {author} {\bibfnamefont {J.}~\bibnamefont {Doyle}}, \bibinfo {author}
  {\bibfnamefont {G.}~\bibnamefont {Gabrielse}}, \bibinfo {author}
  {\bibfnamefont {J.}~\bibnamefont {Haefner}}, \bibinfo {author} {\bibfnamefont
  {N.}~\bibnamefont {Hutzler}}, \bibinfo {author} {\bibfnamefont
  {Z.}~\bibnamefont {Lasner}}, \bibinfo {author} {\bibfnamefont
  {C.}~\bibnamefont {Meisenhelder}}, \bibinfo {author} {\bibfnamefont
  {B.}~\bibnamefont {O'Leary}},  \emph {et~al.},\ }\href {\doibase
  10.1038/s41586-018-0599-8} {\bibfield  {journal} {\bibinfo  {journal}
  {Nature}\ }\textbf {\bibinfo {volume} {562}},\ \bibinfo {pages} {355}
  (\bibinfo {year} {2018})}\BibitemShut {NoStop}%
\bibitem [{\citenamefont {Aggarwal}\ \emph {et~al.}(2018)\citenamefont
  {Aggarwal}, \citenamefont {Bethlem}, \citenamefont {Borschevsky},
  \citenamefont {Denis}, \citenamefont {Esajas}, \citenamefont {Haase},
  \citenamefont {Hao}, \citenamefont {Hoekstra}, \citenamefont {Jungmann},
  \citenamefont {Meijknecht} \emph {et~al.}}]{BaF:2018}%
  \BibitemOpen
  \bibfield  {author} {\bibinfo {author} {\bibfnamefont {P.}~\bibnamefont
  {Aggarwal}}, \bibinfo {author} {\bibfnamefont {H.~L.}\ \bibnamefont
  {Bethlem}}, \bibinfo {author} {\bibfnamefont {A.}~\bibnamefont
  {Borschevsky}}, \bibinfo {author} {\bibfnamefont {M.}~\bibnamefont {Denis}},
  \bibinfo {author} {\bibfnamefont {K.}~\bibnamefont {Esajas}}, \bibinfo
  {author} {\bibfnamefont {P.~A.~B.}\ \bibnamefont {Haase}}, \bibinfo {author}
  {\bibfnamefont {Y.}~\bibnamefont {Hao}}, \bibinfo {author} {\bibfnamefont
  {S.}~\bibnamefont {Hoekstra}}, \bibinfo {author} {\bibfnamefont
  {K.}~\bibnamefont {Jungmann}}, \bibinfo {author} {\bibfnamefont {T.~B.}\
  \bibnamefont {Meijknecht}},  \emph {et~al.},\ }\href {\doibase
  10.1140/epjd/e2018-90192-9} {\bibfield  {journal} {\bibinfo  {journal} {Eur.
  Phys. J. D}\ }\textbf {\bibinfo {volume} {72}},\ \bibinfo {pages} {197}
  (\bibinfo {year} {2018})}\BibitemShut {NoStop}%
\bibitem [{\citenamefont {Boeschoten}\ \emph {et~al.}()\citenamefont
  {Boeschoten}, \citenamefont {Marshall}, \citenamefont {Meijknecht},
  \citenamefont {Touwen}, \citenamefont {Bethlem}, \citenamefont {Borschevsky},
  \citenamefont {Hoekstra}, \citenamefont {van Hofslot}, \citenamefont
  {Jungmann}, \citenamefont {Mooij} \emph {et~al.}}]{boeschoten2023novel}%
  \BibitemOpen
  \bibfield  {author} {\bibinfo {author} {\bibfnamefont {A.}~\bibnamefont
  {Boeschoten}}, \bibinfo {author} {\bibfnamefont {V.}~\bibnamefont
  {Marshall}}, \bibinfo {author} {\bibfnamefont {T.}~\bibnamefont
  {Meijknecht}}, \bibinfo {author} {\bibfnamefont {A.}~\bibnamefont {Touwen}},
  \bibinfo {author} {\bibfnamefont {H.}~\bibnamefont {Bethlem}}, \bibinfo
  {author} {\bibfnamefont {A.}~\bibnamefont {Borschevsky}}, \bibinfo {author}
  {\bibfnamefont {S.}~\bibnamefont {Hoekstra}}, \bibinfo {author}
  {\bibfnamefont {J.}~\bibnamefont {van Hofslot}}, \bibinfo {author}
  {\bibfnamefont {K.}~\bibnamefont {Jungmann}}, \bibinfo {author}
  {\bibfnamefont {M.}~\bibnamefont {Mooij}},  \emph {et~al.},\ }\href {\doibase
  10.48550/arXiv.2303.06402} {\enquote {\bibinfo {title} {arxiv:2303.06402
  [physics.atom-ph] (2023)},}\ }\BibitemShut {NoStop}%
\bibitem [{\citenamefont {Rockenh\"auser}\ \emph {et~al.}(2023)\citenamefont
  {Rockenh\"auser}, \citenamefont {Kogel}, \citenamefont {Pultinevicius},\ and\
  \citenamefont {Langen}}]{PhysRevA.108.062812}%
  \BibitemOpen
  \bibfield  {author} {\bibinfo {author} {\bibfnamefont {M.}~\bibnamefont
  {Rockenh\"auser}}, \bibinfo {author} {\bibfnamefont {F.}~\bibnamefont
  {Kogel}}, \bibinfo {author} {\bibfnamefont {E.}~\bibnamefont
  {Pultinevicius}}, \ and\ \bibinfo {author} {\bibfnamefont {T.}~\bibnamefont
  {Langen}},\ }\href {\doibase 10.1103/PhysRevA.108.062812} {\bibfield
  {journal} {\bibinfo  {journal} {Phys.\ Rev.\ A}\ }\textbf {\bibinfo {volume}
  {108}},\ \bibinfo {pages} {062812} (\bibinfo {year} {2023})}\BibitemShut
  {NoStop}%
\bibitem [{\citenamefont {Kozlov}\ and\ \citenamefont
  {Labzowsky}(1995)}]{KL95}%
  \BibitemOpen
  \bibfield  {author} {\bibinfo {author} {\bibfnamefont {M.}~\bibnamefont
  {Kozlov}}\ and\ \bibinfo {author} {\bibfnamefont {L.}~\bibnamefont
  {Labzowsky}},\ }\href {\doibase 10.1088/0953-4075/28/10/008} {\bibfield
  {journal} {\bibinfo  {journal} {J.\ Phys.\ B}\ }\textbf {\bibinfo {volume}
  {28}},\ \bibinfo {pages} {1933} (\bibinfo {year} {1995})}\BibitemShut
  {NoStop}%
\bibitem [{\citenamefont {Ginges}\ and\ \citenamefont
  {Flambaum}(2004{\natexlab{a}})}]{Ginges:04}%
  \BibitemOpen
  \bibfield  {author} {\bibinfo {author} {\bibfnamefont {J.~S.~M.}\
  \bibnamefont {Ginges}}\ and\ \bibinfo {author} {\bibfnamefont {V.~V.}\
  \bibnamefont {Flambaum}},\ }\href {\doibase 10.1016/j.physrep.2004.03.005}
  {\bibfield  {journal} {\bibinfo  {journal} {Phys.\ Rep.}\ }\textbf {\bibinfo
  {volume} {397}},\ \bibinfo {pages} {63} (\bibinfo {year}
  {2004}{\natexlab{a}})}\BibitemShut {NoStop}%
\bibitem [{\citenamefont {Abbott}\ and\ \citenamefont
  {Sikivie}(1983)}]{abbott1983cosmological}%
  \BibitemOpen
  \bibfield  {author} {\bibinfo {author} {\bibfnamefont {L.~F.}\ \bibnamefont
  {Abbott}}\ and\ \bibinfo {author} {\bibfnamefont {P.}~\bibnamefont
  {Sikivie}},\ }\href {\doibase 10.1016/0370-2693(83)90638-X} {\bibfield
  {journal} {\bibinfo  {journal} {Phys.\ Lett.\ B}\ }\textbf {\bibinfo {volume}
  {120}},\ \bibinfo {pages} {133} (\bibinfo {year} {1983})}\BibitemShut
  {NoStop}%
\bibitem [{\citenamefont {Preskill}\ \emph {et~al.}(1983)\citenamefont
  {Preskill}, \citenamefont {Wise},\ and\ \citenamefont
  {Wilczek}}]{preskill1983cosmology}%
  \BibitemOpen
  \bibfield  {author} {\bibinfo {author} {\bibfnamefont {J.}~\bibnamefont
  {Preskill}}, \bibinfo {author} {\bibfnamefont {M.~B.}\ \bibnamefont {Wise}},
  \ and\ \bibinfo {author} {\bibfnamefont {F.}~\bibnamefont {Wilczek}},\ }\href
  {\doibase 10.1016/0370-2693(83)90637-8} {\bibfield  {journal} {\bibinfo
  {journal} {Phys.\ Lett.\ B}\ }\textbf {\bibinfo {volume} {120}},\ \bibinfo
  {pages} {127} (\bibinfo {year} {1983})}\BibitemShut {NoStop}%
\bibitem [{\citenamefont {Dine}\ and\ \citenamefont
  {Fischler}(1983)}]{dine1983not}%
  \BibitemOpen
  \bibfield  {author} {\bibinfo {author} {\bibfnamefont {M.}~\bibnamefont
  {Dine}}\ and\ \bibinfo {author} {\bibfnamefont {W.}~\bibnamefont
  {Fischler}},\ }\href {\doibase 10.1016/0370-2693(83)90639-1} {\bibfield
  {journal} {\bibinfo  {journal} {Phys.\ Lett.\ B}\ }\textbf {\bibinfo {volume}
  {120}},\ \bibinfo {pages} {137} (\bibinfo {year} {1983})}\BibitemShut
  {NoStop}%
\bibitem [{\citenamefont {Peccei}\ and\ \citenamefont
  {Quinn}(1977)}]{peccei1977cp}%
  \BibitemOpen
  \bibfield  {author} {\bibinfo {author} {\bibfnamefont {R.~D.}\ \bibnamefont
  {Peccei}}\ and\ \bibinfo {author} {\bibfnamefont {H.~R.}\ \bibnamefont
  {Quinn}},\ }\href {\doibase 10.1103/PhysRevLett.38.1440} {\bibfield
  {journal} {\bibinfo  {journal} {Phys. Rev. Lett.}\ }\textbf {\bibinfo
  {volume} {38}},\ \bibinfo {pages} {1440} (\bibinfo {year}
  {1977})}\BibitemShut {NoStop}%
\bibitem [{\citenamefont {Wilczek}(1978)}]{wilczek1978problem}%
  \BibitemOpen
  \bibfield  {author} {\bibinfo {author} {\bibfnamefont {F.}~\bibnamefont
  {Wilczek}},\ }\href {\doibase 10.1103/PhysRevLett.40.279} {\bibfield
  {journal} {\bibinfo  {journal} {Phys.\ Rev.\ Lett.}\ }\textbf {\bibinfo
  {volume} {40}},\ \bibinfo {pages} {279} (\bibinfo {year} {1978})}\BibitemShut
  {NoStop}%
\bibitem [{\citenamefont {Weinberg}(1978)}]{weinberg1978new}%
  \BibitemOpen
  \bibfield  {author} {\bibinfo {author} {\bibfnamefont {S.}~\bibnamefont
  {Weinberg}},\ }\href {\doibase 10.1103/PhysRevLett.40.223} {\bibfield
  {journal} {\bibinfo  {journal} {Phys.\ Rev.\ Lett.}\ }\textbf {\bibinfo
  {volume} {40}},\ \bibinfo {pages} {223} (\bibinfo {year} {1978})}\BibitemShut
  {NoStop}%
\bibitem [{\citenamefont {Svrcek}\ and\ \citenamefont
  {Witten}(2006)}]{svrcek2006axions}%
  \BibitemOpen
  \bibfield  {author} {\bibinfo {author} {\bibfnamefont {P.}~\bibnamefont
  {Svrcek}}\ and\ \bibinfo {author} {\bibfnamefont {E.}~\bibnamefont
  {Witten}},\ }\href {\doibase 10.1088/1126-6708/2006/06/051} {\bibfield
  {journal} {\bibinfo  {journal} {J.\ High\ Energy\ Phys.}\ }\textbf {\bibinfo
  {volume} {2006}},\ \bibinfo {pages} {051} (\bibinfo {year}
  {2006})}\BibitemShut {NoStop}%
\bibitem [{\citenamefont {Arvanitaki}\ \emph {et~al.}(2010)\citenamefont
  {Arvanitaki}, \citenamefont {Dimopoulos}, \citenamefont {Dubovsky},
  \citenamefont {Kaloper},\ and\ \citenamefont
  {March-Russell}}]{arvanitaki2010string}%
  \BibitemOpen
  \bibfield  {author} {\bibinfo {author} {\bibfnamefont {A.}~\bibnamefont
  {Arvanitaki}}, \bibinfo {author} {\bibfnamefont {S.}~\bibnamefont
  {Dimopoulos}}, \bibinfo {author} {\bibfnamefont {S.}~\bibnamefont
  {Dubovsky}}, \bibinfo {author} {\bibfnamefont {N.}~\bibnamefont {Kaloper}}, \
  and\ \bibinfo {author} {\bibfnamefont {J.}~\bibnamefont {March-Russell}},\
  }\href {\doibase 10.1103/PhysRevD.81.123530} {\bibfield  {journal} {\bibinfo
  {journal} {Phys.\ Rev.\ D}\ }\textbf {\bibinfo {volume} {81}},\ \bibinfo
  {pages} {123530} (\bibinfo {year} {2010})}\BibitemShut {NoStop}%
\bibitem [{\citenamefont {O'Hare}\ and\ \citenamefont
  {Vitagliano}(2020)}]{Hare:2020}%
  \BibitemOpen
  \bibfield  {author} {\bibinfo {author} {\bibfnamefont {C.~A.~J.}\
  \bibnamefont {O'Hare}}\ and\ \bibinfo {author} {\bibfnamefont
  {E.}~\bibnamefont {Vitagliano}},\ }\href {\doibase
  10.1103/PhysRevD.102.115026} {\bibfield  {journal} {\bibinfo  {journal}
  {Phys. Rev. D}\ }\textbf {\bibinfo {volume} {102}},\ \bibinfo {pages}
  {115026} (\bibinfo {year} {2020})}\BibitemShut {NoStop}%
\bibitem [{\citenamefont {Agrawal}\ \emph {et~al.}()\citenamefont {Agrawal},
  \citenamefont {Hutzler}, \citenamefont {Kaplan}, \citenamefont {Rajendran},\
  and\ \citenamefont {Reig}}]{agrawal2023searching}%
  \BibitemOpen
  \bibfield  {author} {\bibinfo {author} {\bibfnamefont {P.}~\bibnamefont
  {Agrawal}}, \bibinfo {author} {\bibfnamefont {N.~R.}\ \bibnamefont
  {Hutzler}}, \bibinfo {author} {\bibfnamefont {D.~E.}\ \bibnamefont {Kaplan}},
  \bibinfo {author} {\bibfnamefont {S.}~\bibnamefont {Rajendran}}, \ and\
  \bibinfo {author} {\bibfnamefont {M.}~\bibnamefont {Reig}},\ }\href {\doibase
  10.48550/arXiv.2309.10023} {\enquote {\bibinfo {title} {arxiv:2309.10023
  [hep-ph] (2023)},}\ }\BibitemShut {NoStop}%
\bibitem [{\citenamefont {Kimball}\ \emph {et~al.}(2023)\citenamefont
  {Kimball}, \citenamefont {Budker}, \citenamefont {Chupp}, \citenamefont
  {Geraci}, \citenamefont {Kolkowitz}, \citenamefont {Singh},\ and\
  \citenamefont {Sushkov}}]{kimball2023probing}%
  \BibitemOpen
  \bibfield  {author} {\bibinfo {author} {\bibfnamefont {D.~F.~J.}\
  \bibnamefont {Kimball}}, \bibinfo {author} {\bibfnamefont {D.}~\bibnamefont
  {Budker}}, \bibinfo {author} {\bibfnamefont {T.~E.}\ \bibnamefont {Chupp}},
  \bibinfo {author} {\bibfnamefont {A.~A.}\ \bibnamefont {Geraci}}, \bibinfo
  {author} {\bibfnamefont {S.}~\bibnamefont {Kolkowitz}}, \bibinfo {author}
  {\bibfnamefont {J.~T.}\ \bibnamefont {Singh}}, \ and\ \bibinfo {author}
  {\bibfnamefont {A.~O.}\ \bibnamefont {Sushkov}},\ }\href {\doibase
  10.1103/PhysRevA.108.010101} {\bibfield  {journal} {\bibinfo  {journal}
  {Physical Review A}\ }\textbf {\bibinfo {volume} {108}},\ \bibinfo {pages}
  {010101} (\bibinfo {year} {2023})}\BibitemShut {NoStop}%
\bibitem [{\citenamefont {Stadnik}\ \emph {et~al.}(2018)\citenamefont
  {Stadnik}, \citenamefont {Dzuba},\ and\ \citenamefont
  {Flambaum}}]{Stadnik:2018}%
  \BibitemOpen
  \bibfield  {author} {\bibinfo {author} {\bibfnamefont {Y.~V.}\ \bibnamefont
  {Stadnik}}, \bibinfo {author} {\bibfnamefont {V.~A.}\ \bibnamefont {Dzuba}},
  \ and\ \bibinfo {author} {\bibfnamefont {V.~V.}\ \bibnamefont {Flambaum}},\
  }\href {\doibase 10.1103/PhysRevLett.120.013202} {\bibfield  {journal}
  {\bibinfo  {journal} {Phys.\ Rev.\ Lett.}\ }\textbf {\bibinfo {volume}
  {120}},\ \bibinfo {pages} {013202} (\bibinfo {year} {2018})},\ \bibinfo
  {note} {see also https://arxiv.org/abs/1708.00486v3 (2020)}\BibitemShut
  {NoStop}%
\bibitem [{\citenamefont {Maison}\ \emph
  {et~al.}(2021{\natexlab{a}})\citenamefont {Maison}, \citenamefont {Flambaum},
  \citenamefont {Hutzler},\ and\ \citenamefont {Skripnikov}}]{Maison:2021}%
  \BibitemOpen
  \bibfield  {author} {\bibinfo {author} {\bibfnamefont {D.~E.}\ \bibnamefont
  {Maison}}, \bibinfo {author} {\bibfnamefont {V.~V.}\ \bibnamefont
  {Flambaum}}, \bibinfo {author} {\bibfnamefont {N.~R.}\ \bibnamefont
  {Hutzler}}, \ and\ \bibinfo {author} {\bibfnamefont {L.~V.}\ \bibnamefont
  {Skripnikov}},\ }\href {\doibase 10.1103/PhysRevA.103.022813} {\bibfield
  {journal} {\bibinfo  {journal} {Phys. Rev. A}\ }\textbf {\bibinfo {volume}
  {103}},\ \bibinfo {pages} {022813} (\bibinfo {year}
  {2021}{\natexlab{a}})}\BibitemShut {NoStop}%
\bibitem [{\citenamefont {Maison}\ \emph
  {et~al.}(2021{\natexlab{b}})\citenamefont {Maison}, \citenamefont
  {Skripnikov}, \citenamefont {Oleynichenko},\ and\ \citenamefont
  {Zaitsevskii}}]{maison2021axion}%
  \BibitemOpen
  \bibfield  {author} {\bibinfo {author} {\bibfnamefont {D.~E.}\ \bibnamefont
  {Maison}}, \bibinfo {author} {\bibfnamefont {L.~V.}\ \bibnamefont
  {Skripnikov}}, \bibinfo {author} {\bibfnamefont {A.~V.}\ \bibnamefont
  {Oleynichenko}}, \ and\ \bibinfo {author} {\bibfnamefont {A.~V.}\
  \bibnamefont {Zaitsevskii}},\ }\href {\doibase 10.1063/5.0051590} {\bibfield
  {journal} {\bibinfo  {journal} {J. Chem. Phys.}\ }\textbf {\bibinfo {volume}
  {154}},\ \bibinfo {pages} {224303} (\bibinfo {year}
  {2021}{\natexlab{b}})}\BibitemShut {NoStop}%
\bibitem [{\citenamefont {Maison}\ and\ \citenamefont
  {Skripnikov}(2022)}]{Maison:2022}%
  \BibitemOpen
  \bibfield  {author} {\bibinfo {author} {\bibfnamefont {D.~E.}\ \bibnamefont
  {Maison}}\ and\ \bibinfo {author} {\bibfnamefont {L.~V.}\ \bibnamefont
  {Skripnikov}},\ }\href {\doibase 10.1103/PhysRevA.105.032813} {\bibfield
  {journal} {\bibinfo  {journal} {Phys. Rev. A}\ }\textbf {\bibinfo {volume}
  {105}},\ \bibinfo {pages} {032813} (\bibinfo {year} {2022})}\BibitemShut
  {NoStop}%
\bibitem [{\citenamefont {Prosnyak}\ \emph {et~al.}(2023)\citenamefont
  {Prosnyak}, \citenamefont {Maison},\ and\ \citenamefont
  {Skripnikov}}]{Prosnyak:2023a}%
  \BibitemOpen
  \bibfield  {author} {\bibinfo {author} {\bibfnamefont {S.~D.}\ \bibnamefont
  {Prosnyak}}, \bibinfo {author} {\bibfnamefont {D.~E.}\ \bibnamefont
  {Maison}}, \ and\ \bibinfo {author} {\bibfnamefont {L.~V.}\ \bibnamefont
  {Skripnikov}},\ }\href {\doibase 10.3390/sym15051043} {\bibfield  {journal}
  {\bibinfo  {journal} {Symmetry}\ }\textbf {\bibinfo {volume} {15}} (\bibinfo
  {year} {2023}),\ 10.3390/sym15051043}\BibitemShut {NoStop}%
\bibitem [{\citenamefont {Moody}\ and\ \citenamefont
  {Wilczek}(1984)}]{moody1984new}%
  \BibitemOpen
  \bibfield  {author} {\bibinfo {author} {\bibfnamefont {J.~E.}~\bibnamefont
  {Moody}}\ and\ \bibinfo {author} {\bibfnamefont {F.}~\bibnamefont
  {Wilczek}},\ }\href {\doibase 10.1103/PhysRevD.30.130} {\bibfield  {journal}
  {\bibinfo  {journal} {Phys.\ Rev.\ D}\ }\textbf {\bibinfo {volume} {30}},\
  \bibinfo {pages} {130} (\bibinfo {year} {1984})}\BibitemShut {NoStop}%
\bibitem [{\citenamefont {Dmitriev}\ \emph {et~al.}(1992)\citenamefont
  {Dmitriev}, \citenamefont {Khait}, \citenamefont {Kozlov}, \citenamefont
  {Labzovsky}, \citenamefont {Mitrushenkov}, \citenamefont {Shtoff},\ and\
  \citenamefont {Titov}}]{Dmitriev:92}%
  \BibitemOpen
  \bibfield  {author} {\bibinfo {author} {\bibfnamefont {Y.~Y.}\ \bibnamefont
  {Dmitriev}}, \bibinfo {author} {\bibfnamefont {Y.~G.}\ \bibnamefont {Khait}},
  \bibinfo {author} {\bibfnamefont {M.~G.}\ \bibnamefont {Kozlov}}, \bibinfo
  {author} {\bibfnamefont {L.~N.}\ \bibnamefont {Labzovsky}}, \bibinfo {author}
  {\bibfnamefont {A.~O.}\ \bibnamefont {Mitrushenkov}}, \bibinfo {author}
  {\bibfnamefont {A.~V.}\ \bibnamefont {Shtoff}}, \ and\ \bibinfo {author}
  {\bibfnamefont {A.~V.}\ \bibnamefont {Titov}},\ }\href {\doibase
  10.1016/0375-9601(92)90206-2} {\bibfield  {journal} {\bibinfo  {journal}
  {Phys.\ Lett.\ A}\ }\textbf {\bibinfo {volume} {167}},\ \bibinfo {pages}
  {280} (\bibinfo {year} {1992})}\BibitemShut {NoStop}%
\bibitem [{Note1()}]{Note1}%
  \BibitemOpen
  \bibinfo {note} {It is assumed that the distributions of neutrons and protons
  are equal.}\BibitemShut {Stop}%
\bibitem [{\citenamefont {Visscher}\ and\ \citenamefont
  {Dyall}(1997)}]{Visscher:1997}%
  \BibitemOpen
  \bibfield  {author} {\bibinfo {author} {\bibfnamefont {L.}~\bibnamefont
  {Visscher}}\ and\ \bibinfo {author} {\bibfnamefont {K.~G.}\ \bibnamefont
  {Dyall}},\ }\href {\doibase 10.1006/adnd.1997.0751} {\bibfield  {journal}
  {\bibinfo  {journal} {At. Data Nucl. Data Tables}\ }\textbf {\bibinfo
  {volume} {67}},\ \bibinfo {pages} {207} (\bibinfo {year} {1997})}\BibitemShut
  {NoStop}%
\bibitem [{DIR()}]{DIRAC19}%
  \BibitemOpen
  \href@noop {} {}\bibinfo {note} {DIRAC, a relativistic ab initio electronic
  structure program, Release DIRAC19 (2019), written by A. S. P. Gomes, T.
  Saue, L. Visscher, H. J. Aa. Jensen, and R. Bast, with contributions from I.
  A. Aucar, V. Bakken, K. G. Dyall, S. Dubillard, U. Ekstroem, E. Eliav, T.
  Enevoldsen, E. Fasshauer, T. Fleig, O. Fossgaard, L. Halbert, E. D.
  Hedegaard, T. Helgaker, J. Henriksson, M. Ilias, Ch. R. Jacob, S. Knecht, S.
  Komorovsky, O. Kullie, J. K. Laerdahl, C. V. Larsen, Y. S. Lee, H. S.
  Nataraj, M. K. Nayak, P. Norman, M. Olejniczak, J. Olsen, J. M. H. Olsen, Y.
  C. Park, J. K. Pedersen, M. Pernpointner, R. Di Remigio, K. Ruud, P. Salek,
  B. Schimmelpfennig, B. Senjean, A. Shee, J. Sikkema, A. J. Thorvaldsen, J.
  Thyssen, J. van Stralen, M. L. Vidal, S. Villaume, O. Visser, T. Winther, and
  S. Yamamoto (see http://diracprogram.org).}\BibitemShut {Stop}%
\bibitem [{\citenamefont {Saue}\ \emph {et~al.}(2020)\citenamefont {Saue},
  \citenamefont {Bast}, \citenamefont {Gomes}, \citenamefont {Jensen},
  \citenamefont {Visscher}, \citenamefont {Aucar}, \citenamefont {Di~Remigio},
  \citenamefont {Dyall}, \citenamefont {Eliav}, \citenamefont {Fasshauer},
  \citenamefont {Fleig}, \citenamefont {Halbert}, \citenamefont {Hedegard},
  \citenamefont {Helmich-Paris}, \citenamefont {Ilias}, \citenamefont {Jacob},
  \citenamefont {Knecht}, \citenamefont {Laerdahl}, \citenamefont {Vidal},
  \citenamefont {Nayak}, \citenamefont {Olejniczak}, \citenamefont {Olsen},
  \citenamefont {Pernpointner}, \citenamefont {Senjean}, \citenamefont {Shee},
  \citenamefont {Sunaga},\ and\ \citenamefont {van Stralen}}]{Saue:2020}%
  \BibitemOpen
  \bibfield  {author} {\bibinfo {author} {\bibfnamefont {T.}~\bibnamefont
  {Saue}}, \bibinfo {author} {\bibfnamefont {R.}~\bibnamefont {Bast}}, \bibinfo
  {author} {\bibfnamefont {A.~S.~P.}\ \bibnamefont {Gomes}}, \bibinfo {author}
  {\bibfnamefont {H.~J.~A.}\ \bibnamefont {Jensen}}, \bibinfo {author}
  {\bibfnamefont {L.}~\bibnamefont {Visscher}}, \bibinfo {author}
  {\bibfnamefont {I.~A.}\ \bibnamefont {Aucar}}, \bibinfo {author}
  {\bibfnamefont {R.}~\bibnamefont {Di~Remigio}}, \bibinfo {author}
  {\bibfnamefont {K.~G.}\ \bibnamefont {Dyall}}, \bibinfo {author}
  {\bibfnamefont {E.}~\bibnamefont {Eliav}}, \bibinfo {author} {\bibfnamefont
  {E.}~\bibnamefont {Fasshauer}}, \bibinfo {author} {\bibfnamefont
  {T.}~\bibnamefont {Fleig}}, \bibinfo {author} {\bibfnamefont
  {L.}~\bibnamefont {Halbert}}, \bibinfo {author} {\bibfnamefont {E.~D.}\
  \bibnamefont {Hedegard}}, \bibinfo {author} {\bibfnamefont {B.}~\bibnamefont
  {Helmich-Paris}}, \bibinfo {author} {\bibfnamefont {M.}~\bibnamefont
  {Ilias}}, \bibinfo {author} {\bibfnamefont {C.~R.}\ \bibnamefont {Jacob}},
  \bibinfo {author} {\bibfnamefont {S.}~\bibnamefont {Knecht}}, \bibinfo
  {author} {\bibfnamefont {J.~K.}\ \bibnamefont {Laerdahl}}, \bibinfo {author}
  {\bibfnamefont {M.~L.}\ \bibnamefont {Vidal}}, \bibinfo {author}
  {\bibfnamefont {M.~K.}\ \bibnamefont {Nayak}}, \bibinfo {author}
  {\bibfnamefont {M.}~\bibnamefont {Olejniczak}}, \bibinfo {author}
  {\bibfnamefont {J.~M.~H.}\ \bibnamefont {Olsen}}, \bibinfo {author}
  {\bibfnamefont {M.}~\bibnamefont {Pernpointner}}, \bibinfo {author}
  {\bibfnamefont {B.}~\bibnamefont {Senjean}}, \bibinfo {author} {\bibfnamefont
  {A.}~\bibnamefont {Shee}}, \bibinfo {author} {\bibfnamefont {A.}~\bibnamefont
  {Sunaga}}, \ and\ \bibinfo {author} {\bibfnamefont {J.~N.~P.}\ \bibnamefont
  {van Stralen}},\ }\href {\doibase 10.1063/5.0004844} {\bibfield  {journal}
  {\bibinfo  {journal} {J.\ Chem.\ Phys.}\ }\textbf {\bibinfo {volume} {152}},\
  \bibinfo {pages} {204104} (\bibinfo {year} {2020})}\BibitemShut {NoStop}%
\bibitem [{MRC()}]{MRCC2020}%
  \BibitemOpen
  \href@noop {} {\enquote {\bibinfo {title} {{{\sc mrcc}}},}\ }\bibinfo {note}
  {M. K\'{a}llay, P. R. Nagy, D. Mester, Z. Rolik, G. Samu, J. Csontos, J.
  Cs\'{o}ka, P. B. Szab\'{o}, L. Gyevi-Nagy, B. H\'{e}gely, I. Ladj\'{a}nszki,
  L. Szegedy, B. Lad\'{o}czki, K. Petrov, M. Farkas, P. D. Mezei, and \'{a}.
  Ganyecz: The {\sc mrcc} program system: Accurate quantum chemistry from water
  to proteins, J. Chem. Phys. 152, 074107 (2020); {\sc mrcc}, a quantum
  chemical program suite written by M. K\'{a}llay, P. R. Nagy, D. Mester, Z.
  Rolik, G. Samu, J. Csontos, J. Cs\'{o}ka, P. B. Szab\'{o}, L. Gyevi-Nagy, B.
  H\'{e}gely, I. Ladj\'{a}nszki, L. Szegedy, B. Lad\'{o}czki, K. Petrov, M.
  Farkas, P. D. Mezei, and \'{A}. Ganyecz. See www.mrcc.hu.}\BibitemShut
  {Stop}%
\bibitem [{\citenamefont {K\'{a}llay}\ and\ \citenamefont
  {Surj\'{a}n}(2001)}]{Kallay:1}%
  \BibitemOpen
  \bibfield  {author} {\bibinfo {author} {\bibfnamefont {M.}~\bibnamefont
  {K\'{a}llay}}\ and\ \bibinfo {author} {\bibfnamefont {P.~R.}\ \bibnamefont
  {Surj\'{a}n}},\ }\href {\doibase 10.1063/1.1383290} {\bibfield  {journal}
  {\bibinfo  {journal} {J.\ Chem.\ Phys.}\ }\textbf {\bibinfo {volume} {115}},\
  \bibinfo {pages} {2945} (\bibinfo {year} {2001})}\BibitemShut {NoStop}%
\bibitem [{\citenamefont {K\'{a}llay}\ \emph {et~al.}(2002)\citenamefont
  {K\'{a}llay}, \citenamefont {Szalay},\ and\ \citenamefont
  {Surj\'{a}n}}]{Kallay:2}%
  \BibitemOpen
  \bibfield  {author} {\bibinfo {author} {\bibfnamefont {M.}~\bibnamefont
  {K\'{a}llay}}, \bibinfo {author} {\bibfnamefont {P.~G.}\ \bibnamefont
  {Szalay}}, \ and\ \bibinfo {author} {\bibfnamefont {P.~R.}\ \bibnamefont
  {Surj\'{a}n}},\ }\href {\doibase 10.1063/1.1483856} {\bibfield  {journal}
  {\bibinfo  {journal} {J.\ Chem.\ Phys.}\ }\textbf {\bibinfo {volume} {117}},\
  \bibinfo {pages} {980} (\bibinfo {year} {2002})}\BibitemShut {NoStop}%
\bibitem [{\citenamefont {Visscher}\ \emph {et~al.}(1996)\citenamefont
  {Visscher}, \citenamefont {Lee},\ and\ \citenamefont {Dyall}}]{Visscher:96a}%
  \BibitemOpen
  \bibfield  {author} {\bibinfo {author} {\bibfnamefont {L.}~\bibnamefont
  {Visscher}}, \bibinfo {author} {\bibfnamefont {T.~J.}\ \bibnamefont {Lee}}, \
  and\ \bibinfo {author} {\bibfnamefont {K.~G.}\ \bibnamefont {Dyall}},\ }\href
  {\doibase 10.1063/1.472655} {\bibfield  {journal} {\bibinfo  {journal} {J.\
  Chem.\ Phys.}\ }\textbf {\bibinfo {volume} {105}},\ \bibinfo {pages} {8769}
  (\bibinfo {year} {1996})}\BibitemShut {NoStop}%
\bibitem [{\citenamefont {Bartlett}\ and\ \citenamefont
  {Musia{\l}}(2007)}]{Bartlett:2007}%
  \BibitemOpen
  \bibfield  {author} {\bibinfo {author} {\bibfnamefont {R.~J.}\ \bibnamefont
  {Bartlett}}\ and\ \bibinfo {author} {\bibfnamefont {M.}~\bibnamefont
  {Musia{\l}}},\ }\href {\doibase 10.1103/RevModPhys.79.291} {\bibfield
  {journal} {\bibinfo  {journal} {Rev. Mod. Phys.}\ }\textbf {\bibinfo {volume}
  {79}},\ \bibinfo {pages} {291} (\bibinfo {year} {2007})}\BibitemShut
  {NoStop}%
\bibitem [{\citenamefont {Huber}\ and\ \citenamefont
  {Herzberg}(1979)}]{Huber:1979}%
  \BibitemOpen
  \bibfield  {author} {\bibinfo {author} {\bibfnamefont {K.~P.}\ \bibnamefont
  {Huber}}\ and\ \bibinfo {author} {\bibfnamefont {G.}~\bibnamefont
  {Herzberg}},\ }\href {\doibase 10.1007/978-1-4757-0961-2_2} {\emph {\bibinfo
  {title} {Constants of Diatomic Molecules}}}\ (\bibinfo  {publisher} {Van
  Nostrand-Reinhold},\ \bibinfo {address} {New York},\ \bibinfo {year}
  {1979})\BibitemShut {NoStop}%
\bibitem [{\citenamefont {Ryzlewicz}\ and\ \citenamefont
  {T{\"o}rring}(1980)}]{Ryzlewicz:1980}%
  \BibitemOpen
  \bibfield  {author} {\bibinfo {author} {\bibfnamefont {C.}~\bibnamefont
  {Ryzlewicz}}\ and\ \bibinfo {author} {\bibfnamefont {T.}~\bibnamefont
  {T{\"o}rring}},\ }\href {\doibase
  https://doi.org/10.1016/0301-0104(80)80107-8} {\bibfield  {journal} {\bibinfo
   {journal} {Chem. Phys.}\ }\textbf {\bibinfo {volume} {51}},\ \bibinfo
  {pages} {329} (\bibinfo {year} {1980})}\BibitemShut {NoStop}%
\bibitem [{\citenamefont {Skripnikov}\ \emph {et~al.}(2017)\citenamefont
  {Skripnikov}, \citenamefont {Maison},\ and\ \citenamefont
  {Mosyagin}}]{Skripnikov:17a}%
  \BibitemOpen
  \bibfield  {author} {\bibinfo {author} {\bibfnamefont {L.~V.}\ \bibnamefont
  {Skripnikov}}, \bibinfo {author} {\bibfnamefont {D.~E.}\ \bibnamefont
  {Maison}}, \ and\ \bibinfo {author} {\bibfnamefont {N.~S.}\ \bibnamefont
  {Mosyagin}},\ }\href {\doibase 10.1103/PhysRevA.95.022507} {\bibfield
  {journal} {\bibinfo  {journal} {Phys.\ Rev.\ A}\ }\textbf {\bibinfo {volume}
  {95}},\ \bibinfo {pages} {022507} (\bibinfo {year} {2017})}\BibitemShut
  {NoStop}%
\bibitem [{\citenamefont {Skripnikov}\ and\ \citenamefont
  {Titov}(2015)}]{Skripnikov:15b}%
  \BibitemOpen
  \bibfield  {author} {\bibinfo {author} {\bibfnamefont {L.~V.}\ \bibnamefont
  {Skripnikov}}\ and\ \bibinfo {author} {\bibfnamefont {A.~V.}\ \bibnamefont
  {Titov}},\ }\href {\doibase 10.1103/PhysRevA.91.042504} {\bibfield  {journal}
  {\bibinfo  {journal} {Phys. Rev. A}\ }\textbf {\bibinfo {volume} {91}},\
  \bibinfo {pages} {042504} (\bibinfo {year} {2015})}\BibitemShut {NoStop}%
\bibitem [{\citenamefont {Dyall}(2016)}]{Dyall:2016}%
  \BibitemOpen
  \bibfield  {author} {\bibinfo {author} {\bibfnamefont {K.~G.}\ \bibnamefont
  {Dyall}},\ }\href {\doibase 10.1007/s00214-016-1884-y} {\bibfield  {journal}
  {\bibinfo  {journal} {Theor. Chem. Acc.}\ }\textbf {\bibinfo {volume}
  {135}},\ \bibinfo {pages} {128} (\bibinfo {year} {2016})}\BibitemShut
  {NoStop}%
\bibitem [{\citenamefont {Dyall}(2009)}]{dyall2009relativistic}%
  \BibitemOpen
  \bibfield  {author} {\bibinfo {author} {\bibfnamefont {K.~G.}\ \bibnamefont
  {Dyall}},\ }\href {\doibase 10.1021/jp905057q} {\bibfield  {journal}
  {\bibinfo  {journal} {J.\ Phys.\ Chem.\ A}\ }\textbf {\bibinfo {volume}
  {113}},\ \bibinfo {pages} {12638} (\bibinfo {year} {2009})}\BibitemShut
  {NoStop}%
\bibitem [{\citenamefont {Dyall}(2012)}]{Dyall:12}%
  \BibitemOpen
  \bibfield  {author} {\bibinfo {author} {\bibfnamefont {K.~G.}\ \bibnamefont
  {Dyall}},\ }\href {\doibase 10.1007/s00214-012-1217-8} {\bibfield  {journal}
  {\bibinfo  {journal} {Theor. Chem. Acc.}\ }\textbf {\bibinfo {volume}
  {131}},\ \bibinfo {pages} {1217} (\bibinfo {year} {2012})}\BibitemShut
  {NoStop}%
\bibitem [{\citenamefont {Johnson}\ and\ \citenamefont
  {Soff}(1985)}]{Johnson:1985}%
  \BibitemOpen
  \bibfield  {author} {\bibinfo {author} {\bibfnamefont {W.}~\bibnamefont
  {Johnson}}\ and\ \bibinfo {author} {\bibfnamefont {G.}~\bibnamefont {Soff}},\
  }\href {\doibase https://doi.org/10.1016/0092-640X(85)90010-5} {\bibfield
  {journal} {\bibinfo  {journal} {At. Data Nucl. Data Tables}\ }\textbf
  {\bibinfo {volume} {33}},\ \bibinfo {pages} {405} (\bibinfo {year}
  {1985})}\BibitemShut {NoStop}%
\bibitem [{\citenamefont {Haase}\ \emph {et~al.}(2021)\citenamefont {Haase},
  \citenamefont {Doeglas}, \citenamefont {Boeschoten}, \citenamefont {Eliav},
  \citenamefont {Iliaš}, \citenamefont {Aggarwal}, \citenamefont {Bethlem},
  \citenamefont {Borschevsky}, \citenamefont {Esajas}, \citenamefont {Hao},
  \citenamefont {Hoekstra}, \citenamefont {Marshall}, \citenamefont
  {Meijknecht}, \citenamefont {Mooij}, \citenamefont {Steinebach},
  \citenamefont {Timmermans}, \citenamefont {Touwen}, \citenamefont {Ubachs},
  \citenamefont {Willmann}, \citenamefont {Yin},\ and\ \citenamefont {eEDM
  Collaboration)}}]{BaF_enhancement_2021}%
  \BibitemOpen
  \bibfield  {author} {\bibinfo {author} {\bibfnamefont {P.~A.~B.}\
  \bibnamefont {Haase}}, \bibinfo {author} {\bibfnamefont {D.~J.}\ \bibnamefont
  {Doeglas}}, \bibinfo {author} {\bibfnamefont {A.}~\bibnamefont {Boeschoten}},
  \bibinfo {author} {\bibfnamefont {E.}~\bibnamefont {Eliav}}, \bibinfo
  {author} {\bibfnamefont {M.}~\bibnamefont {Iliaš}}, \bibinfo {author}
  {\bibfnamefont {P.}~\bibnamefont {Aggarwal}}, \bibinfo {author}
  {\bibfnamefont {H.~L.}\ \bibnamefont {Bethlem}}, \bibinfo {author}
  {\bibfnamefont {A.}~\bibnamefont {Borschevsky}}, \bibinfo {author}
  {\bibfnamefont {K.}~\bibnamefont {Esajas}}, \bibinfo {author} {\bibfnamefont
  {Y.}~\bibnamefont {Hao}}, \bibinfo {author} {\bibfnamefont {S.}~\bibnamefont
  {Hoekstra}}, \bibinfo {author} {\bibfnamefont {V.~R.}\ \bibnamefont
  {Marshall}}, \bibinfo {author} {\bibfnamefont {T.~B.}\ \bibnamefont
  {Meijknecht}}, \bibinfo {author} {\bibfnamefont {M.~C.}\ \bibnamefont
  {Mooij}}, \bibinfo {author} {\bibfnamefont {K.}~\bibnamefont {Steinebach}},
  \bibinfo {author} {\bibfnamefont {R.~G.~E.}\ \bibnamefont {Timmermans}},
  \bibinfo {author} {\bibfnamefont {A.~P.}\ \bibnamefont {Touwen}}, \bibinfo
  {author} {\bibfnamefont {W.}~\bibnamefont {Ubachs}}, \bibinfo {author}
  {\bibfnamefont {L.}~\bibnamefont {Willmann}}, \bibinfo {author}
  {\bibfnamefont {Y.}~\bibnamefont {Yin}}, \ and\ \bibinfo {author}
  {\bibfnamefont {N.}~\bibnamefont {eEDM Collaboration)}},\ }\href {\doibase
  10.1063/5.0047344} {\bibfield  {journal} {\bibinfo  {journal} {J.\ Chem.\
  Phys.}\ }\textbf {\bibinfo {volume} {155}},\ \bibinfo {pages} {034309}
  (\bibinfo {year} {2021})}\BibitemShut {NoStop}%
\bibitem [{\citenamefont {Ginges}\ and\ \citenamefont
  {Flambaum}(2004{\natexlab{b}})}]{GFreview}%
  \BibitemOpen
  \bibfield  {author} {\bibinfo {author} {\bibfnamefont {J.~S.~M.}\
  \bibnamefont {Ginges}}\ and\ \bibinfo {author} {\bibfnamefont {V.~V.}\
  \bibnamefont {Flambaum}},\ }\href {\doibase 10.1016/j.physrep.2004.03.005}
  {\bibfield  {journal} {\bibinfo  {journal} {Phys.\ Rep.}\ }\textbf {\bibinfo
  {volume} {397}},\ \bibinfo {pages} {63} (\bibinfo {year}
  {2004}{\natexlab{b}})}\BibitemShut {NoStop}%
\bibitem [{\citenamefont {Jung}(2013)}]{Jung:13}%
  \BibitemOpen
  \bibfield  {author} {\bibinfo {author} {\bibfnamefont {M.}~\bibnamefont
  {Jung}},\ }\href {\doibase 10.1007/JHEP05(2013)168} {\bibfield  {journal}
  {\bibinfo  {journal} {J. High Energy Phys.}\ }\textbf {\bibinfo {volume}
  {2013}},\ \bibinfo {pages} {168} (\bibinfo {year} {2013})}\BibitemShut
  {NoStop}%
\bibitem [{\citenamefont {Heckel}\ \emph {et~al.}(2008)\citenamefont {Heckel},
  \citenamefont {Adelberger}, \citenamefont {Cramer}, \citenamefont {Cook},
  \citenamefont {Schlamminger},\ and\ \citenamefont {Schmidt}}]{Heckel:2008}%
  \BibitemOpen
  \bibfield  {author} {\bibinfo {author} {\bibfnamefont {B.~R.}\ \bibnamefont
  {Heckel}}, \bibinfo {author} {\bibfnamefont {E.~G.}\ \bibnamefont
  {Adelberger}}, \bibinfo {author} {\bibfnamefont {C.~E.}\ \bibnamefont
  {Cramer}}, \bibinfo {author} {\bibfnamefont {T.~S.}\ \bibnamefont {Cook}},
  \bibinfo {author} {\bibfnamefont {S.}~\bibnamefont {Schlamminger}}, \ and\
  \bibinfo {author} {\bibfnamefont {U.}~\bibnamefont {Schmidt}},\ }\href
  {\doibase 10.1103/PhysRevD.78.092006} {\bibfield  {journal} {\bibinfo
  {journal} {Phys. Rev. D}\ }\textbf {\bibinfo {volume} {78}},\ \bibinfo
  {pages} {092006} (\bibinfo {year} {2008})}\BibitemShut {NoStop}%
\bibitem [{\citenamefont {Wineland}\ \emph {et~al.}(1991)\citenamefont
  {Wineland}, \citenamefont {Bollinger}, \citenamefont {Heinzen}, \citenamefont
  {Itano},\ and\ \citenamefont {Raizen}}]{Wineland:1991}%
  \BibitemOpen
  \bibfield  {author} {\bibinfo {author} {\bibfnamefont {D.~J.}\ \bibnamefont
  {Wineland}}, \bibinfo {author} {\bibfnamefont {J.~J.}\ \bibnamefont
  {Bollinger}}, \bibinfo {author} {\bibfnamefont {D.~J.}\ \bibnamefont
  {Heinzen}}, \bibinfo {author} {\bibfnamefont {W.~M.}\ \bibnamefont {Itano}},
  \ and\ \bibinfo {author} {\bibfnamefont {M.~G.}\ \bibnamefont {Raizen}},\
  }\href {\doibase 10.1103/PhysRevLett.67.1735} {\bibfield  {journal} {\bibinfo
   {journal} {Phys. Rev. Lett.}\ }\textbf {\bibinfo {volume} {67}},\ \bibinfo
  {pages} {1735} (\bibinfo {year} {1991})}\BibitemShut {NoStop}%
\bibitem [{\citenamefont {Lee}\ \emph {et~al.}(2018)\citenamefont {Lee},
  \citenamefont {Almasi},\ and\ \citenamefont {Romalis}}]{Lee:2018}%
  \BibitemOpen
  \bibfield  {author} {\bibinfo {author} {\bibfnamefont {J.}~\bibnamefont
  {Lee}}, \bibinfo {author} {\bibfnamefont {A.}~\bibnamefont {Almasi}}, \ and\
  \bibinfo {author} {\bibfnamefont {M.}~\bibnamefont {Romalis}},\ }\href
  {\doibase 10.1103/PhysRevLett.120.161801} {\bibfield  {journal} {\bibinfo
  {journal} {Phys. Rev. Lett.}\ }\textbf {\bibinfo {volume} {120}},\ \bibinfo
  {pages} {161801} (\bibinfo {year} {2018})}\BibitemShut {NoStop}%
\bibitem [{\citenamefont {Hoedl}\ \emph {et~al.}(2011)\citenamefont {Hoedl},
  \citenamefont {Fleischer}, \citenamefont {Adelberger},\ and\ \citenamefont
  {Heckel}}]{hoedl2011improved}%
  \BibitemOpen
  \bibfield  {author} {\bibinfo {author} {\bibfnamefont {S.~A.}~\bibnamefont
  {Hoedl}}, \bibinfo {author} {\bibfnamefont {S.~M.}~\bibnamefont {Fleischer}},
  \bibinfo {author} {\bibfnamefont {E.~G.}~\bibnamefont {Adelberger}}, \ and\
  \bibinfo {author} {\bibfnamefont {B.~R.}~\bibnamefont {Heckel}},\ }\href
  {\doibase 10.1103/PhysRevLett.106.041801} {\bibfield  {journal} {\bibinfo
  {journal} {Phys.\ Rev.\ Lett.}\ }\textbf {\bibinfo {volume} {106}},\ \bibinfo
  {pages} {041801} (\bibinfo {year} {2011})}\BibitemShut {NoStop}%
\bibitem [{\citenamefont {Aprile}\ \emph {et~al.}(2019)\citenamefont {Aprile},
  \citenamefont {Aalbers}, \citenamefont {Agostini}, \citenamefont {Alfonsi},
  \citenamefont {Althueser}, \citenamefont {Amaro}, \citenamefont {Antochi},
  \citenamefont {Angelino}, \citenamefont {Arneodo}, \citenamefont {Barge},
  \citenamefont {Baudis}, \citenamefont {Bauermeister}, \citenamefont
  {Bellagamba}, \citenamefont {Benabderrahmane}, \citenamefont {Berger},
  \citenamefont {Breur}, \citenamefont {Brown}, \citenamefont {Brown},
  \citenamefont {Bruenner}, \citenamefont {Bruno}, \citenamefont {Budnik},
  \citenamefont {Capelli}, \citenamefont {Cardoso}, \citenamefont {Cichon},
  \citenamefont {Coderre}, \citenamefont {Colijn}, \citenamefont {Conrad},
  \citenamefont {Cussonneau}, \citenamefont {Decowski}, \citenamefont
  {de~Perio}, \citenamefont {Depoian}, \citenamefont {Di~Gangi}, \citenamefont
  {Di~Giovanni}, \citenamefont {Diglio}, \citenamefont {Elykov}, \citenamefont
  {Eurin}, \citenamefont {Fei}, \citenamefont {Ferella}, \citenamefont
  {Fieguth}, \citenamefont {Fulgione}, \citenamefont {Gaemers}, \citenamefont
  {Gallo~Rosso}, \citenamefont {Galloway}, \citenamefont {Gao}, \citenamefont
  {Garbini}, \citenamefont {Grandi}, \citenamefont {Greene}, \citenamefont
  {Hasterok}, \citenamefont {Hils}, \citenamefont {Hogenbirk}, \citenamefont
  {Howlett}, \citenamefont {Iacovacci}, \citenamefont {Itay}, \citenamefont
  {Joerg}, \citenamefont {Kazama}, \citenamefont {Kish}, \citenamefont
  {Kobayashi}, \citenamefont {Koltman}, \citenamefont {Kopec}, \citenamefont
  {Landsman}, \citenamefont {Lang}, \citenamefont {Levinson}, \citenamefont
  {Lin}, \citenamefont {Lindemann}, \citenamefont {Lindner}, \citenamefont
  {Lombardi}, \citenamefont {Lopes}, \citenamefont {L\'opez~Fune},
  \citenamefont {Macolino}, \citenamefont {Mahlstedt}, \citenamefont
  {Manfredini}, \citenamefont {Marignetti}, \citenamefont
  {Marrod\'an~Undagoitia}, \citenamefont {Masbou}, \citenamefont {Mastroianni},
  \citenamefont {Messina}, \citenamefont {Micheneau}, \citenamefont {Miller},
  \citenamefont {Molinario}, \citenamefont {Mor\aa{}}, \citenamefont
  {Mosbacher}, \citenamefont {Murra}, \citenamefont {Naganoma}, \citenamefont
  {Ni}, \citenamefont {Oberlack}, \citenamefont {Odgers}, \citenamefont
  {Palacio}, \citenamefont {Pelssers}, \citenamefont {Peres}, \citenamefont
  {Pienaar}, \citenamefont {Pizzella}, \citenamefont {Plante}, \citenamefont
  {Podviianiuk}, \citenamefont {Qin}, \citenamefont {Qiu}, \citenamefont
  {Ram\'{\i}rez~Garc\'{\i}a}, \citenamefont {Reichard}, \citenamefont {Riedel},
  \citenamefont {Rocchetti}, \citenamefont {Rupp}, \citenamefont {dos Santos},
  \citenamefont {Sartorelli}, \citenamefont {\ifmmode \check{S}\else
  \v{S}\fi{}ar\ifmmode \check{c}\else \v{c}\fi{}evi\ifmmode~\acute{c}\else
  \'{c}\fi{}}, \citenamefont {Scheibelhut}, \citenamefont {Schindler},
  \citenamefont {Schreiner}, \citenamefont {Schulte}, \citenamefont {Schumann},
  \citenamefont {Scotto~Lavina}, \citenamefont {Selvi}, \citenamefont {Shagin},
  \citenamefont {Shockley}, \citenamefont {Silva}, \citenamefont {Simgen},
  \citenamefont {Therreau}, \citenamefont {Thers}, \citenamefont {Toschi},
  \citenamefont {Trinchero}, \citenamefont {Tunnell}, \citenamefont {Upole},
  \citenamefont {Vargas}, \citenamefont {Volta}, \citenamefont {Wack},
  \citenamefont {Wang}, \citenamefont {Wei}, \citenamefont {Weinheimer},
  \citenamefont {Wenz}, \citenamefont {Wittweg}, \citenamefont {Wulf},
  \citenamefont {Ye}, \citenamefont {Zhang}, \citenamefont {Zhu},\ and\
  \citenamefont {Zopounidis}}]{XENON1T:2019}%
  \BibitemOpen
  \bibfield  {author} {\bibinfo {author} {\bibfnamefont {E.}~\bibnamefont
  {Aprile}}, \bibinfo {author} {\bibfnamefont {J.}~\bibnamefont {Aalbers}},
  \bibinfo {author} {\bibfnamefont {F.}~\bibnamefont {Agostini}}, \bibinfo
  {author} {\bibfnamefont {M.}~\bibnamefont {Alfonsi}}, \bibinfo {author}
  {\bibfnamefont {L.}~\bibnamefont {Althueser}}, \bibinfo {author}
  {\bibfnamefont {F.~D.}\ \bibnamefont {Amaro}}, \bibinfo {author}
  {\bibfnamefont {V.~C.}\ \bibnamefont {Antochi}}, \bibinfo {author}
  {\bibfnamefont {E.}~\bibnamefont {Angelino}}, \bibinfo {author}
  {\bibfnamefont {F.}~\bibnamefont {Arneodo}}, \bibinfo {author} {\bibfnamefont
  {D.}~\bibnamefont {Barge}}, \bibinfo {author} {\bibfnamefont
  {L.}~\bibnamefont {Baudis}}, \bibinfo {author} {\bibfnamefont
  {B.}~\bibnamefont {Bauermeister}}, \bibinfo {author} {\bibfnamefont
  {L.}~\bibnamefont {Bellagamba}}, \bibinfo {author} {\bibfnamefont {M.~L.}\
  \bibnamefont {Benabderrahmane}}, \bibinfo {author} {\bibfnamefont
  {T.}~\bibnamefont {Berger}}, \bibinfo {author} {\bibfnamefont {P.~A.}\
  \bibnamefont {Breur}}, \bibinfo {author} {\bibfnamefont {A.}~\bibnamefont
  {Brown}}, \bibinfo {author} {\bibfnamefont {E.}~\bibnamefont {Brown}},
  \bibinfo {author} {\bibfnamefont {S.}~\bibnamefont {Bruenner}}, \bibinfo
  {author} {\bibfnamefont {G.}~\bibnamefont {Bruno}}, \bibinfo {author}
  {\bibfnamefont {R.}~\bibnamefont {Budnik}}, \bibinfo {author} {\bibfnamefont
  {C.}~\bibnamefont {Capelli}}, \bibinfo {author} {\bibfnamefont {J.~M.~R.}\
  \bibnamefont {Cardoso}}, \bibinfo {author} {\bibfnamefont {D.}~\bibnamefont
  {Cichon}}, \bibinfo {author} {\bibfnamefont {D.}~\bibnamefont {Coderre}},
  \bibinfo {author} {\bibfnamefont {A.~P.}\ \bibnamefont {Colijn}}, \bibinfo
  {author} {\bibfnamefont {J.}~\bibnamefont {Conrad}}, \bibinfo {author}
  {\bibfnamefont {J.~P.}\ \bibnamefont {Cussonneau}}, \bibinfo {author}
  {\bibfnamefont {M.~P.}\ \bibnamefont {Decowski}}, \bibinfo {author}
  {\bibfnamefont {P.}~\bibnamefont {de~Perio}}, \bibinfo {author}
  {\bibfnamefont {A.}~\bibnamefont {Depoian}}, \bibinfo {author} {\bibfnamefont
  {P.}~\bibnamefont {Di~Gangi}}, \bibinfo {author} {\bibfnamefont
  {A.}~\bibnamefont {Di~Giovanni}}, \bibinfo {author} {\bibfnamefont
  {S.}~\bibnamefont {Diglio}}, \bibinfo {author} {\bibfnamefont
  {A.}~\bibnamefont {Elykov}}, \bibinfo {author} {\bibfnamefont
  {G.}~\bibnamefont {Eurin}}, \bibinfo {author} {\bibfnamefont
  {J.}~\bibnamefont {Fei}}, \bibinfo {author} {\bibfnamefont {A.~D.}\
  \bibnamefont {Ferella}}, \bibinfo {author} {\bibfnamefont {A.}~\bibnamefont
  {Fieguth}}, \bibinfo {author} {\bibfnamefont {W.}~\bibnamefont {Fulgione}},
  \bibinfo {author} {\bibfnamefont {P.}~\bibnamefont {Gaemers}}, \bibinfo
  {author} {\bibfnamefont {A.}~\bibnamefont {Gallo~Rosso}}, \bibinfo {author}
  {\bibfnamefont {M.}~\bibnamefont {Galloway}}, \bibinfo {author}
  {\bibfnamefont {F.}~\bibnamefont {Gao}}, \bibinfo {author} {\bibfnamefont
  {M.}~\bibnamefont {Garbini}}, \bibinfo {author} {\bibfnamefont
  {L.}~\bibnamefont {Grandi}}, \bibinfo {author} {\bibfnamefont
  {Z.}~\bibnamefont {Greene}}, \bibinfo {author} {\bibfnamefont
  {C.}~\bibnamefont {Hasterok}}, \bibinfo {author} {\bibfnamefont
  {C.}~\bibnamefont {Hils}}, \bibinfo {author} {\bibfnamefont {E.}~\bibnamefont
  {Hogenbirk}}, \bibinfo {author} {\bibfnamefont {J.}~\bibnamefont {Howlett}},
  \bibinfo {author} {\bibfnamefont {M.}~\bibnamefont {Iacovacci}}, \bibinfo
  {author} {\bibfnamefont {R.}~\bibnamefont {Itay}}, \bibinfo {author}
  {\bibfnamefont {F.}~\bibnamefont {Joerg}}, \bibinfo {author} {\bibfnamefont
  {S.}~\bibnamefont {Kazama}}, \bibinfo {author} {\bibfnamefont
  {A.}~\bibnamefont {Kish}}, \bibinfo {author} {\bibfnamefont {M.}~\bibnamefont
  {Kobayashi}}, \bibinfo {author} {\bibfnamefont {G.}~\bibnamefont {Koltman}},
  \bibinfo {author} {\bibfnamefont {A.}~\bibnamefont {Kopec}}, \bibinfo
  {author} {\bibfnamefont {H.}~\bibnamefont {Landsman}}, \bibinfo {author}
  {\bibfnamefont {R.~F.}\ \bibnamefont {Lang}}, \bibinfo {author}
  {\bibfnamefont {L.}~\bibnamefont {Levinson}}, \bibinfo {author}
  {\bibfnamefont {Q.}~\bibnamefont {Lin}}, \bibinfo {author} {\bibfnamefont
  {S.}~\bibnamefont {Lindemann}}, \bibinfo {author} {\bibfnamefont
  {M.}~\bibnamefont {Lindner}}, \bibinfo {author} {\bibfnamefont
  {F.}~\bibnamefont {Lombardi}}, \bibinfo {author} {\bibfnamefont {J.~A.~M.}\
  \bibnamefont {Lopes}}, \bibinfo {author} {\bibfnamefont {E.}~\bibnamefont
  {L\'opez~Fune}}, \bibinfo {author} {\bibfnamefont {C.}~\bibnamefont
  {Macolino}}, \bibinfo {author} {\bibfnamefont {J.}~\bibnamefont {Mahlstedt}},
  \bibinfo {author} {\bibfnamefont {A.}~\bibnamefont {Manfredini}}, \bibinfo
  {author} {\bibfnamefont {F.}~\bibnamefont {Marignetti}}, \bibinfo {author}
  {\bibfnamefont {T.}~\bibnamefont {Marrod\'an~Undagoitia}}, \bibinfo {author}
  {\bibfnamefont {J.}~\bibnamefont {Masbou}}, \bibinfo {author} {\bibfnamefont
  {S.}~\bibnamefont {Mastroianni}}, \bibinfo {author} {\bibfnamefont
  {M.}~\bibnamefont {Messina}}, \bibinfo {author} {\bibfnamefont
  {K.}~\bibnamefont {Micheneau}}, \bibinfo {author} {\bibfnamefont
  {K.}~\bibnamefont {Miller}}, \bibinfo {author} {\bibfnamefont
  {A.}~\bibnamefont {Molinario}}, \bibinfo {author} {\bibfnamefont
  {K.}~\bibnamefont {Mor\aa{}}}, \bibinfo {author} {\bibfnamefont
  {Y.}~\bibnamefont {Mosbacher}}, \bibinfo {author} {\bibfnamefont
  {M.}~\bibnamefont {Murra}}, \bibinfo {author} {\bibfnamefont
  {J.}~\bibnamefont {Naganoma}}, \bibinfo {author} {\bibfnamefont
  {K.}~\bibnamefont {Ni}}, \bibinfo {author} {\bibfnamefont {U.}~\bibnamefont
  {Oberlack}}, \bibinfo {author} {\bibfnamefont {K.}~\bibnamefont {Odgers}},
  \bibinfo {author} {\bibfnamefont {J.}~\bibnamefont {Palacio}}, \bibinfo
  {author} {\bibfnamefont {B.}~\bibnamefont {Pelssers}}, \bibinfo {author}
  {\bibfnamefont {R.}~\bibnamefont {Peres}}, \bibinfo {author} {\bibfnamefont
  {J.}~\bibnamefont {Pienaar}}, \bibinfo {author} {\bibfnamefont
  {V.}~\bibnamefont {Pizzella}}, \bibinfo {author} {\bibfnamefont
  {G.}~\bibnamefont {Plante}}, \bibinfo {author} {\bibfnamefont
  {R.}~\bibnamefont {Podviianiuk}}, \bibinfo {author} {\bibfnamefont
  {J.}~\bibnamefont {Qin}}, \bibinfo {author} {\bibfnamefont {H.}~\bibnamefont
  {Qiu}}, \bibinfo {author} {\bibfnamefont {D.}~\bibnamefont
  {Ram\'{\i}rez~Garc\'{\i}a}}, \bibinfo {author} {\bibfnamefont
  {S.}~\bibnamefont {Reichard}}, \bibinfo {author} {\bibfnamefont
  {B.}~\bibnamefont {Riedel}}, \bibinfo {author} {\bibfnamefont
  {A.}~\bibnamefont {Rocchetti}}, \bibinfo {author} {\bibfnamefont
  {N.}~\bibnamefont {Rupp}}, \bibinfo {author} {\bibfnamefont {J.~M.~F.}\
  \bibnamefont {dos Santos}}, \bibinfo {author} {\bibfnamefont
  {G.}~\bibnamefont {Sartorelli}}, \bibinfo {author} {\bibfnamefont
  {N.}~\bibnamefont {\ifmmode \check{S}\else \v{S}\fi{}ar\ifmmode
  \check{c}\else \v{c}\fi{}evi\ifmmode~\acute{c}\else \'{c}\fi{}}}, \bibinfo
  {author} {\bibfnamefont {M.}~\bibnamefont {Scheibelhut}}, \bibinfo {author}
  {\bibfnamefont {S.}~\bibnamefont {Schindler}}, \bibinfo {author}
  {\bibfnamefont {J.}~\bibnamefont {Schreiner}}, \bibinfo {author}
  {\bibfnamefont {D.}~\bibnamefont {Schulte}}, \bibinfo {author} {\bibfnamefont
  {M.}~\bibnamefont {Schumann}}, \bibinfo {author} {\bibfnamefont
  {L.}~\bibnamefont {Scotto~Lavina}}, \bibinfo {author} {\bibfnamefont
  {M.}~\bibnamefont {Selvi}}, \bibinfo {author} {\bibfnamefont
  {P.}~\bibnamefont {Shagin}}, \bibinfo {author} {\bibfnamefont
  {E.}~\bibnamefont {Shockley}}, \bibinfo {author} {\bibfnamefont
  {M.}~\bibnamefont {Silva}}, \bibinfo {author} {\bibfnamefont
  {H.}~\bibnamefont {Simgen}}, \bibinfo {author} {\bibfnamefont
  {C.}~\bibnamefont {Therreau}}, \bibinfo {author} {\bibfnamefont
  {D.}~\bibnamefont {Thers}}, \bibinfo {author} {\bibfnamefont
  {F.}~\bibnamefont {Toschi}}, \bibinfo {author} {\bibfnamefont
  {G.}~\bibnamefont {Trinchero}}, \bibinfo {author} {\bibfnamefont
  {C.}~\bibnamefont {Tunnell}}, \bibinfo {author} {\bibfnamefont
  {N.}~\bibnamefont {Upole}}, \bibinfo {author} {\bibfnamefont
  {M.}~\bibnamefont {Vargas}}, \bibinfo {author} {\bibfnamefont
  {G.}~\bibnamefont {Volta}}, \bibinfo {author} {\bibfnamefont
  {O.}~\bibnamefont {Wack}}, \bibinfo {author} {\bibfnamefont {H.}~\bibnamefont
  {Wang}}, \bibinfo {author} {\bibfnamefont {Y.}~\bibnamefont {Wei}}, \bibinfo
  {author} {\bibfnamefont {C.}~\bibnamefont {Weinheimer}}, \bibinfo {author}
  {\bibfnamefont {D.}~\bibnamefont {Wenz}}, \bibinfo {author} {\bibfnamefont
  {C.}~\bibnamefont {Wittweg}}, \bibinfo {author} {\bibfnamefont
  {J.}~\bibnamefont {Wulf}}, \bibinfo {author} {\bibfnamefont {J.}~\bibnamefont
  {Ye}}, \bibinfo {author} {\bibfnamefont {Y.}~\bibnamefont {Zhang}}, \bibinfo
  {author} {\bibfnamefont {T.}~\bibnamefont {Zhu}}, \ and\ \bibinfo {author}
  {\bibfnamefont {J.~P.}\ \bibnamefont {Zopounidis}} (\bibinfo {collaboration}
  {XENON Collaboration}),\ }\href {\doibase 10.1103/PhysRevLett.123.251801}
  {\bibfield  {journal} {\bibinfo  {journal} {Phys. Rev. Lett.}\ }\textbf
  {\bibinfo {volume} {123}},\ \bibinfo {pages} {251801} (\bibinfo {year}
  {2019})}\BibitemShut {NoStop}%
\bibitem [{\citenamefont {Youdin}\ \emph {et~al.}(1996)\citenamefont {Youdin},
  \citenamefont {Krause~Jr.}, \citenamefont {Jagannathan}, \citenamefont
  {Hunter},\ and\ \citenamefont {Lamoreaux}}]{youdin1996limits}%
  \BibitemOpen
  \bibfield  {author} {\bibinfo {author} {\bibfnamefont {A.~N.}\ \bibnamefont
  {Youdin}}, \bibinfo {author} {\bibfnamefont {D.}~\bibnamefont {Krause~Jr.}},
  \bibinfo {author} {\bibfnamefont {K.}~\bibnamefont {Jagannathan}}, \bibinfo
  {author} {\bibfnamefont {L.~R.}\ \bibnamefont {Hunter}}, \ and\ \bibinfo
  {author} {\bibfnamefont {S.~K.}\ \bibnamefont {Lamoreaux}},\ }\href {\doibase
  10.1103/PhysRevLett.77.2170} {\bibfield  {journal} {\bibinfo  {journal}
  {Phys.\ Rev.\ Lett.}\ }\textbf {\bibinfo {volume} {77}},\ \bibinfo {pages}
  {2170} (\bibinfo {year} {1996})}\BibitemShut {NoStop}%
\bibitem [{\citenamefont {Ni}\ \emph {et~al.}(1999)\citenamefont {Ni},
  \citenamefont {Pan}, \citenamefont {Yeh}, \citenamefont {Hou},\ and\
  \citenamefont {Wan}}]{ni1999search}%
  \BibitemOpen
  \bibfield  {author} {\bibinfo {author} {\bibfnamefont {W.-T.}\ \bibnamefont
  {Ni}}, \bibinfo {author} {\bibfnamefont {S.~S.}\ \bibnamefont {Pan}},
  \bibinfo {author} {\bibfnamefont {H.-C.}\ \bibnamefont {Yeh}}, \bibinfo
  {author} {\bibfnamefont {L.-S.}\ \bibnamefont {Hou}}, \ and\ \bibinfo
  {author} {\bibfnamefont {J.}~\bibnamefont {Wan}},\ }\href {\doibase
  10.1103/PhysRevLett.82.2439} {\bibfield  {journal} {\bibinfo  {journal}
  {Phys.\ Rev.\ Lett.}\ }\textbf {\bibinfo {volume} {82}},\ \bibinfo {pages}
  {2439} (\bibinfo {year} {1999})}\BibitemShut {NoStop}%
\bibitem [{\citenamefont {Hammond}\ \emph {et~al.}(2007)\citenamefont
  {Hammond}, \citenamefont {Speake}, \citenamefont {Trenkel},\ and\
  \citenamefont {Pat{\'o}n}}]{hammond2007new}%
  \BibitemOpen
  \bibfield  {author} {\bibinfo {author} {\bibfnamefont {G.~D.}\ \bibnamefont
  {Hammond}}, \bibinfo {author} {\bibfnamefont {C.~C.}\ \bibnamefont {Speake}},
  \bibinfo {author} {\bibfnamefont {C.}~\bibnamefont {Trenkel}}, \ and\
  \bibinfo {author} {\bibfnamefont {A.~P.}\ \bibnamefont {Pat{\'o}n}},\ }\href
  {\doibase 10.1103/PhysRevLett.98.081101} {\bibfield  {journal} {\bibinfo
  {journal} {Phys.\ Rev.\ Lett.}\ }\textbf {\bibinfo {volume} {98}},\ \bibinfo
  {pages} {081101} (\bibinfo {year} {2007})}\BibitemShut {NoStop}%
\bibitem [{\citenamefont {Terrano}\ \emph {et~al.}(2015)\citenamefont
  {Terrano}, \citenamefont {Adelberger}, \citenamefont {Lee},\ and\
  \citenamefont {Heckel}}]{terrano2015short}%
  \BibitemOpen
  \bibfield  {author} {\bibinfo {author} {\bibfnamefont {W.~A.}~\bibnamefont
  {Terrano}}, \bibinfo {author} {\bibfnamefont {E.~G.}~\bibnamefont {Adelberger}},
  \bibinfo {author} {\bibfnamefont {J.~G.}~\bibnamefont {Lee}}, \ and\ \bibinfo
  {author} {\bibfnamefont {B.~R.}~\bibnamefont {Heckel}},\ }\href@noop {}
  {\bibfield  {journal} {\bibinfo  {journal} {Phys.\ Rev.\ Lett.}\ }\textbf
  {\bibinfo {volume} {115}},\ \bibinfo {pages} {201801} (\bibinfo {year}
  {2015})}\BibitemShut {NoStop}%
\bibitem [{\citenamefont {Crescini}\ \emph {et~al.}(2017)\citenamefont
  {Crescini}, \citenamefont {Braggio}, \citenamefont {Carugno}, \citenamefont
  {Falferi}, \citenamefont {Ortolan},\ and\ \citenamefont
  {Ruoso}}]{crescini2017improved}%
  \BibitemOpen
  \bibfield  {author} {\bibinfo {author} {\bibfnamefont {N.}~\bibnamefont
  {Crescini}}, \bibinfo {author} {\bibfnamefont {C.}~\bibnamefont {Braggio}},
  \bibinfo {author} {\bibfnamefont {G.}~\bibnamefont {Carugno}}, \bibinfo
  {author} {\bibfnamefont {P.}~\bibnamefont {Falferi}}, \bibinfo {author}
  {\bibfnamefont {A.}~\bibnamefont {Ortolan}}, \ and\ \bibinfo {author}
  {\bibfnamefont {G.}~\bibnamefont {Ruoso}},\ }\href@noop {} {\bibfield
  {journal} {\bibinfo  {journal} {Phys.\ Lett.\ B}\ }\textbf {\bibinfo {volume}
  {773}},\ \bibinfo {pages} {677} (\bibinfo {year} {2017})}\BibitemShut
  {NoStop}%
\bibitem [{\citenamefont {Rong}\ \emph {et~al.}(2018)\citenamefont {Rong},
  \citenamefont {Wang}, \citenamefont {Geng}, \citenamefont {Qin},
  \citenamefont {Guo}, \citenamefont {Jiao}, \citenamefont {Xie}, \citenamefont
  {Wang}, \citenamefont {Huang}, \citenamefont {Shi} \emph
  {et~al.}}]{rong2018searching}%
  \BibitemOpen
  \bibfield  {author} {\bibinfo {author} {\bibfnamefont {X.}~\bibnamefont
  {Rong}}, \bibinfo {author} {\bibfnamefont {M.}~\bibnamefont {Wang}}, \bibinfo
  {author} {\bibfnamefont {J.}~\bibnamefont {Geng}}, \bibinfo {author}
  {\bibfnamefont {X.}~\bibnamefont {Qin}}, \bibinfo {author} {\bibfnamefont
  {M.}~\bibnamefont {Guo}}, \bibinfo {author} {\bibfnamefont {M.}~\bibnamefont
  {Jiao}}, \bibinfo {author} {\bibfnamefont {Y.}~\bibnamefont {Xie}}, \bibinfo
  {author} {\bibfnamefont {P.}~\bibnamefont {Wang}}, \bibinfo {author}
  {\bibfnamefont {P.}~\bibnamefont {Huang}}, \bibinfo {author} {\bibfnamefont
  {F.}~\bibnamefont {Shi}},  \emph {et~al.},\ }\href@noop {} {\bibfield
  {journal} {\bibinfo  {journal} {Nature communications}\ }\textbf {\bibinfo
  {volume} {9}},\ \bibinfo {pages} {739} (\bibinfo {year} {2018})}\BibitemShut
  {NoStop}%
\end{thebibliography}
\end{document}